\PassOptionsToPackage{table}{xcolor}
\documentclass{article}
\usepackage{iclr2027_conference,times}
\iclrfinalcopy
\usepackage[T1]{fontenc}
\usepackage{amsmath,amssymb,amsfonts,booktabs,microtype}
\usepackage{enumitem,array,xspace,graphicx,etoolbox,placeins,float}
\usepackage{xcolor}
\definecolor{finegrain}{HTML}{F0F4FA}
\definecolor{coarsegrain}{HTML}{FAF3E8}
\definecolor{mesograin}{HTML}{E8F4EF}
\definecolor{atomgeometry}{HTML}{F2ECF8}
\usepackage{multirow}
\usepackage{wrapfig}
\usepackage{needspace}
\AtBeginEnvironment{wrapfigure}{\setlength{\intextsep}{5pt}\setlength{\abovecaptionskip}{4pt}}
\AtBeginEnvironment{wraptable}{\setlength{\intextsep}{5pt}\setlength{\abovecaptionskip}{4pt}}
\newcommand{\finishwrap}{\par
  \ifnum\value{WF@wrappedlines}>1
    \vspace{\dimexpr\value{WF@wrappedlines}\baselineskip-2\baselineskip\relax}
  \fi\WFclear}
\usepackage{hyperref}
\makeatletter
\newif\if@capstart
\apptocmd{\@floatboxreset}{%
  \phantomsection
  \global\let\hc@currentHref\@currentHref
  \global\@capstarttrue
}{}{\PackageError{anchorpose}{Could not install float-top link anchors}{}}
\makeatother
\hypersetup{colorlinks=true,citecolor=blue,linkcolor=black,urlcolor=blue,
  pdftitle={AnchorPose for Geometry-Aware MOF Assembly through Meso-Grained Pose Generation},
  pdfauthor={Zhonglong Peng; Rui Jiao; Chang Chen; Geng Zhong; Qiuliang Liu; Shifeng Jin}}
\setlist{nosep,leftmargin=*}
\newcommand{\method}{\textsc{AnchorPose}\xspace}

\newcommand{\R}{\mathbb{R}}
\newcommand{\SO}{\mathrm{SO}}
\newcommand{\stol}{\texttt{stol}}
\title{\fontsize{16}{20}\selectfont AnchorPose for Geometry-Aware MOF Assembly\\through Meso-Grained Pose Generation}
\author{\normalfont
\begin{minipage}{\dimexpr\textwidth-2\tabcolsep\relax}
\centering
Zhonglong Peng\textsuperscript{1,2,*}\quad Rui Jiao\textsuperscript{3,*}\quad
Chang Chen\textsuperscript{1,2}\\[3pt]
Geng Zhong\textsuperscript{1,2}\quad Qiuliang Liu\textsuperscript{1,2}\quad
Shifeng Jin\textsuperscript{1,2}\\[6pt]
{\small \textsuperscript{1}Institute of Physics, Chinese Academy of Sciences\\
\textsuperscript{2}University of the Chinese Academy of Sciences\\
\textsuperscript{3}Tsinghua University}
\end{minipage}}
\date{}
\begin{document}
\raggedbottom
\maketitle
\fancyhead[L]{Preprint}

\begin{abstract}
Predicting metal--organic framework (MOF) structures from given building blocks requires recovering their positions and orientations in a periodic crystal.
The spatial effects of rotation errors are geometry-dependent and anisotropic. The same angular error can produce different atomic displacements depending on block size, shape, and rotation axis.
Angular error alone, without reference to the specific block geometry, therefore cannot fully describe the spatial consequences of a pose error.
We introduce \method, a meso-grained pose generation framework that incorporates this geometric dependence into its generative representation.
It represents each block through a small set of representative atoms, combines their local geometry with the current spatial state, and generates their coordinates with Bayesian Flow Networks.
Known atom correspondences enable rigid alignment to recover complete building-block poses and return geometrically consistent points to the generation process.
This design connects point-level spatial prediction with block-level structural constraints. Geometry participates in the pose state and its prediction, while rigid reconstruction preserves intra-block structure without treating all atomic coordinates as assembly variables.
On the MOF benchmark, \method improves single-candidate match rates over the compared block-level and all-atom baselines.
\end{abstract}

\section{Introduction}
Metal--organic frameworks (MOFs) are porous crystalline materials in which organic linkers connect metal nodes into periodic networks.
The choice of building blocks tunes pore dimensions and chemistry, making MOFs candidates for gas storage, separation, and catalysis~\citep{furukawa2013mof}.
Computational screening uses three-dimensional structures to evaluate pore accessibility and adsorption environments~\citep{furukawa2013mof,kim2026atommof}.

We study MOF structure prediction from the molecular graphs and multiplicities of the constituent building blocks, with the lattice and complete atomic arrangement as outputs.
This task requires placing and orienting the blocks within a periodic crystal.
Building blocks vary in size and shape. Accurate assembly must account for how their specific geometries affect the spatial arrangement.

Existing methods primarily operate at either the block or atom level.
Block-level methods, such as MOFFlow and MOF-BFN, describe each block's overall pose through translation and rotation~\citep{kim2024mofflow,jiao2025mofbfn}.
This representation is compact and preserves intra-block geometry through rigid transformations, but the shape information used for pose prediction is typically compressed into block-level features.
These features provide shape information to the predictor, but the pose state and its noise remain defined in a rotation space independently of the specific building-block geometry.
Atom-level methods, such as AtomMOF, directly generate all atomic coordinates, explicitly representing spatial geometry and accommodating structural flexibility, but also introducing more coordinate degrees of freedom~\citep{kim2026atommof}.
These two approaches motivate a representation that preserves the overall structure of each block while directly generating the spatial positions of its different parts, without predicting every atomic coordinate.

The pose of a building block describes its overall placement in space, as expressed by the positions of its constituent atoms~\citep{bregier2017pose}.
Given a local conformation, rotation and translation fully specify this pose, but angular differences alone cannot fully describe its spatial error.
The same angular error produces different atomic displacements depending on block size, shape, and rotation axis.
Even within one block, atoms farther from the rotation axis move more.
This geometry dependence motivates a representation in which changes to the generated variables directly describe the spatial displacement of specific parts of a block.
Supplying shape features to a rotation predictor provides geometric context. Generating corresponding atom positions also makes these spatial locations the quantities the network learns to recover.
A small set of representative atoms can serve this role, while the known local conformation connects their predicted positions to the pose of the complete block.
Geometry then enters the generation process through the current point state, point-level prediction and supervision, and rigid reconstruction.

We introduce \method, a geometry-aware framework for meso-grained pose generation.
\method selects a small set of spatially dispersed representative atoms from each block as anchors and combines their local geometry with the current spatial state to generate anchor positions using Bayesian Flow Networks~\citep{graves2023bfn}.
Rigid alignment then recovers the pose of the complete building block~\citep{kabsch1976}.
The same anchors organize the noisy state, clean-coordinate prediction, spatial supervision, and rigidly reconstructed feedback.
Rigid reconstruction recovers all atomic positions from the known local conformation while preserving each block's internal geometry.

Our main contributions are as follows.
\begin{itemize}
\item We analyze how building-block geometry affects the atomic displacements caused by rotation errors in MOF assembly. This analysis motivates incorporating geometry into the pose representation itself, beyond its role as a conditioning feature.
\item We introduce \method, a meso-grained generator that carries a small set of fixed-identity anchors through noisy states, point prediction, coordinate supervision, and rigidly reconstructed feedback.
\item We improve single-candidate match rates and five-candidate loose match rate over the compared methods, while matching AtomMOF-L's five-candidate strict performance.
\end{itemize}

\begin{figure}[t]
\centering
\includegraphics[width=\textwidth]{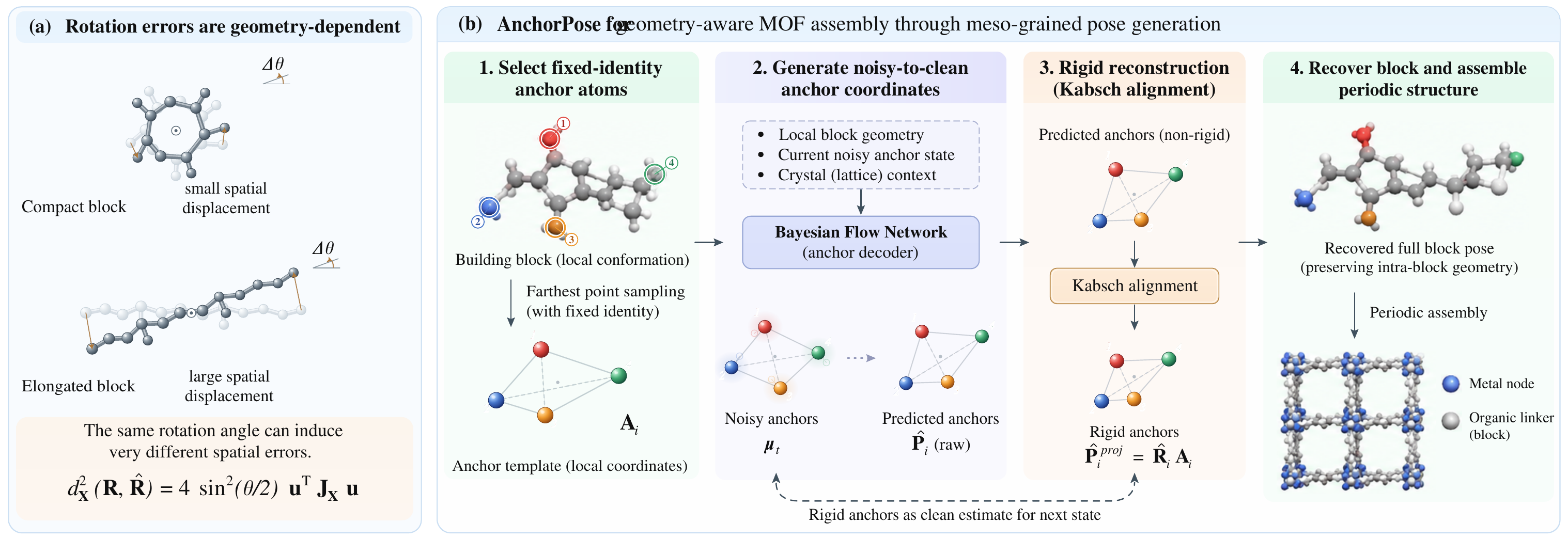}
\caption{\textbf{Overview of AnchorPose.} \textbf{(a)} The same rotation angle can lead to very different spatial displacements for building blocks with different geometries. \textbf{(b)} AnchorPose represents each building block by a small set of fixed-identity anchor atoms, generates their coordinates with a Bayesian Flow Network conditioned on local geometry and crystal context, and recovers the full block pose via Kabsch alignment. The rigidly reconstructed anchors are fed back as the clean estimate for the next generation step, enabling geometry-aware and structurally consistent MOF assembly.}
\label{fig:overview}
\end{figure}

\section{Related work}
\paragraph{Representations for MOF structure generation.}
Rigid-block representations reduce the number of generative variables and preserve intra-block geometry during MOF assembly.
MOFDiff generates coarse-grained block identities and positions, then assembles an all-atom MOF structure~\citep{fu2024mofdiff}.
For structure prediction from given building blocks, MOFFlow and MOF-BFN generate block translations, rotations, and lattice parameters~\citep{kim2024mofflow,jiao2025mofbfn}.
To model internal flexibility, MOFFlow-2 introduces linker torsions~\citep{kim2025mofflow2}.
AtomMOF conditions all-atom coordinate generation on local geometry and chemical information, allowing bond lengths, bond angles, and coordination geometry to vary~\citep{kim2026atommof}.
AnchorPose generates the coordinates of a few representative atoms with fixed identities (anchors), then recovers complete blocks by rigidly aligning prepared local conformations to those anchors.
This meso-grained representation learns spatial positions directly while preserving intra-block geometry during assembly.

\paragraph{Geometry and pose modeling.}
Object geometry determines the spatial consequences of rotation error.
Studies of rigid-object pose distances and molecular rigid-motion root-mean-square deviation measure pose differences through point displacement, relating spatial error to object size, shape, and rotation axis~\citep{bregier2017pose,popov2014rigidrmsd}.
Analysis of rotational alignment in point-cloud diffusion further connects the rotation posterior to point-cloud geometry, showing how geometry affects orientation uncertainty under noisy observations~\citep{daigavane2025denoiser}.
These relations motivate our atom-based pose representation.

Deep Closest Point learns correspondences between two observed point clouds, then recovers their rigid transformation with differentiable singular value decomposition~\citep{wang2019dcp}.
Molecular docking methods also learn point positions to recover poses.
EquiBind predicts paired ligand and receptor keypoints under binding-region supervision, then aligns them to recover the ligand's rotation and translation~\citep{stark2022equibind}.
These keypoints act as binding-interface landmarks that determine the relative placement of the two molecules.
Its flexible variant also predicts a deformed ligand point cloud and fits rotatable-bond torsions to that cloud while preserving local bond lengths and angles~\citep{stark2022equibind}.
For generative docking, DiffDock models ligand poses through diffusion over translations, rotations, and torsion angles~\citep{corso2023diffdock}.
GeoDiff generates molecular conformations by equivariantly diffusing atomic coordinates~\citep{xu2022geodiff}.

AnchorPose's anchors represent each block's shape and pose.
We select spatially dispersed atoms with fixed identities, so point prediction requires no correspondence estimation.
These atoms need not lie at inter-block connections.
Their coordinates define noisy states, prediction targets, and coordinate supervision. Rigidly recovered points feed subsequent generation steps.
Both methods use point alignment. EquiBind's keypoints mark docking interfaces, while AnchorPose's anchors express block geometry and pose throughout generation.

\section{Geometry-aware pose representation}\label{sec:geometry}

\subsection{Problem setup}\label{sec:setup}

We study conditional MOF structure generation from the molecular graphs and multiplicities of metal nodes and organic linkers. These inputs specify the composition of the crystal, but leave its lattice, block placements, and inter-block connectivity unknown. For each block instance $i$, we prepare a local conformation $\mathbf{X}_i=\{\mathbf{x}_{ia}\}_{a=1}^{n_i}$ from its molecular graph and center it at its full-atom centroid. The assembly model then generates a lattice $\mathbf{L}$, a fractional center $\mathbf{f}_i\in[0,1)^3$, and a rotation $\mathbf{R}_i\in\mathrm{SO}(3)$, which together place each atom at
\begin{equation}
\mathbf{y}_{ia}=\mathbf{c}_i+\mathbf{R}_i\mathbf{x}_{ia},\qquad \mathbf{c}_i=\mathbf{L}\mathbf{f}_i,
\label{eq:reconstruct}
\end{equation}

Here $\mathbf{c}_i$ is the Cartesian block center, and the columns of $\mathbf{L}$ are the lattice vectors. The local conformations remain fixed during assembly, so the model determines how the blocks are arranged within a periodic cell. Atom identities and intra-block molecular graphs are given. Inter-block coordination edges and connection labels are not. AnchorPose represents each block's orientation through the center-relative coordinates of selected atoms and recovers its rotation by rigid alignment.

\begin{figure}[t]
\centering
\raisebox{-0.5\height}{\includegraphics[width=0.47\textwidth]{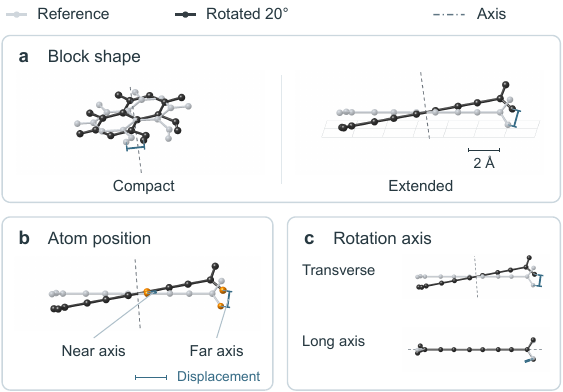}}\hfill
\raisebox{-0.5\height}{\includegraphics[width=0.49\textwidth]{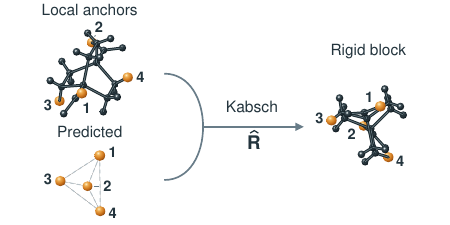}}
\caption{\textbf{From geometry-dependent pose error to rigid recovery.} \textbf{Left.} Reference and $20^\circ$-rotated poses show how atomic displacement depends on block shape, distance from the axis, and rotation axis. Blue markers show projected displacements. \textbf{Right.} Kabsch aligns local anchors with predicted points and rotates the complete block. Numbers identify corresponding atoms. Translation is omitted.}
\label{fig:motivation}\label{fig:anchorrecovery}
\end{figure}

\subsection{Geometry of pose error}\label{sec:spatialerror}

The choice of pose representation matters because MOF reconstruction is evaluated through atomic positions. An angular error alone does not specify the resulting displacement. Rotating a long linker across its long axis moves its ends much farther than rotating it about that axis. The same angle therefore produces different spatial errors across blocks and rotation axes (Figure~\ref{fig:motivation}).

For two orientations $\mathbf{R}_i$ and $\widehat{\mathbf{R}}_i$ at a fixed block center, the mean squared atomic displacement is
\begin{equation}
\begin{aligned}
d_{\mathbf{X}_i}^2(\mathbf{R}_i,\widehat{\mathbf{R}}_i)
&=\frac{1}{n_i}\sum_a\|\widehat{\mathbf{R}}_i\mathbf{x}_{ia}-\mathbf{R}_i\mathbf{x}_{ia}\|^2
=4\sin^2(\theta/2)\,\mathbf{u}^\top \mathbf{J}_{\mathbf{X}_i}\mathbf{u},\\
\mathbf{J}_{\mathbf{X}_i}
&=\frac{1}{n_i}\sum_a\left(\|\mathbf{x}_{ia}\|^2\mathbf{I}-\mathbf{x}_{ia}\mathbf{x}_{ia}^\top\right),
\end{aligned}
\label{eq:spatialmetric}
\end{equation}

Here $\theta$ and $\mathbf{u}$ are the angle and local unit axis of $\mathbf{R}_i^\top\widehat{\mathbf{R}}_i$. The factor $\mathbf{u}^\top \mathbf{J}_{\mathbf{X}_i}\mathbf{u}$ measures the mean squared distance of the atoms from that axis. This equal-atom-weight form of established pose distances~\citep{popov2014rigidrmsd,bregier2017pose} separates angular error from the block's size and directional sensitivity. Appendix~\ref{app:theory} derives the identity.

A common angular perturbation thus has no common spatial scale across blocks. We use anchor coordinates to express uncertainty in the spatial domain where reconstruction errors occur. This choice concerns the uncertainty model. A rotation fully represents orientation, but its spatial effect depends on the block geometry. Geometry-dependent rotational noise offers an alternative, which we evaluate alongside anchor-coordinate states in Section~\ref{sec:eval}.

\subsection{Anchor coordinates and rigid recovery}\label{sec:pointrepresentation}

For each block, we use farthest-point sampling to select $K_i=\min(4,n_i)$ spatially dispersed atoms as anchors. We keep the same anchor atoms throughout training and generation. At orientation $\mathbf{R}_i$, their local and rotated coordinates satisfy
\begin{equation}
\mathbf{A}_i=[\mathbf{a}_{i1},\ldots,\mathbf{a}_{iK_i}]\in\mathbb R^{3\times K_i},\qquad \mathbf{P}_i=\mathbf{R}_i\mathbf{A}_i.
\label{eq:anchorrepresentation}
\end{equation}

Both coordinate sets are relative to the full-block centroid, whose position we model separately. Each generated point retains its correspondence to a local atom throughout the flow.
For two orientations of the same block,
\begin{equation}
\frac{1}{K_i}\|\widehat{\mathbf{P}}_i-\mathbf{P}_i\|_F^2
=\frac{1}{K_i}\sum_k\|(\widehat{\mathbf{R}}_i-\mathbf{R}_i)\mathbf{a}_{ik}\|^2.
\label{eq:pointmetric}
\end{equation}

This error measures displacement over the selected atoms and depends on their distances from the rotation axis. By generating their coordinates, we capture this geometric dependence with only a few points per block. The anchor error need not equal the full-atom error. We examine how anchor selection affects geometric coverage in Appendix~\ref{app:representation}.

\Needspace{8\baselineskip}
Noisy states and point predictions can distort the anchor layout (Figure~\ref{fig:anchorrecovery}). We recover a rigid block by fitting local anchors to $\widehat{\mathbf{P}}_i^{\mathrm{raw}}$ with centered Kabsch alignment~\citep{kabsch1976} by solving
\begin{equation}
\widehat{\mathbf{R}}_i^{0}
=\underset{\mathbf{R}\in\mathrm{SO}(3)}{\arg\min}
\sum_k\left\|
\mathbf{R}(\mathbf{a}_{ik}-\bar{\mathbf{a}}_i)
-(\widehat{\mathbf{p}}_{ik}^{\mathrm{raw}}-\bar{\widehat{\mathbf{p}}}_i^{\mathrm{raw}})
\right\|^2.
\label{eq:anchorprojection}
\end{equation}

The bars denote anchor centroids. We apply the initial rotation $\widehat{\mathbf{R}}_i^0$ to all atoms in $\mathbf{X}_i$ and translate the block to its generated center, preserving its conformation even when predicted anchors are distorted. Appendix~\ref{app:config} describes the undetermined rotations of collinear and single-point templates.

\paragraph{Proposition 1 (geometry-dependent orientation uncertainty).}
Consider non-collinear anchors observed as $\mathbf{z}_k=\mathbf{R}^*\mathbf{a}_k+\tau\boldsymbol{\epsilon}_k$, with independent $\boldsymbol{\epsilon}_k\sim\mathcal N(\mathbf{0},\mathbf{I})$.
Let $\widehat{\mathbf{R}}=\mathbf{R}^*\exp([\boldsymbol{\delta\omega}]_\times)$ be their centered Kabsch estimate, where $[\cdot]_\times$ denotes the cross-product matrix.
For small noise, the local rotation error has covariance
\begin{equation}
\operatorname{Cov}(\boldsymbol{\delta\omega})\simeq\tau^2\mathbf{J}_{\mathbf{A}}^{-1},
\qquad
\mathbf{J}_{\mathbf{A}}=\sum_k\left(\|\bar{\mathbf{a}}_k\|^2\mathbf{I}-\bar{\mathbf{a}}_k\bar{\mathbf{a}}_k^\top\right),
\label{eq:anchoruncertainty}
\end{equation}
where $\bar{\mathbf{a}}_k=\mathbf{a}_k-K^{-1}\sum_l \mathbf{a}_l$.
Appendix~\ref{app:pointuncertainty} derives this first-order approximation.

The same positional noise therefore constrains orientation more tightly when anchors extend farther from a rotation axis.
For a fixed shape, scaling the anchors by $r$ reduces the local angular covariance by $r^2$. Near-collinearity instead leaves a weakly constrained axis.
Unlike $\mathbf{J}_{\mathbf{X}}$, which measures full-atom displacement, $\mathbf{J}_{\mathbf{A}}$ describes the selected anchors about their own centroid.
Section~\ref{sec:main_diagnostics} tests this size dependence empirically.

\FloatBarrier
\section{AnchorPose}\label{sec:method}\label{sec:assembly}

We first prepare local block conformations and derive their training targets by aligning them to the corresponding crystal blocks (\textsection~\ref{sec:fit}). Given these local conformations, we use a Bayesian Flow Network (BFN) to jointly generate the lattice, block centers, and anchor coordinates, predicting clean coordinates from the current state at each step (\textsection~\ref{sec:anchors}). We then recover rigid blocks by Kabsch alignment and apply two pose updates to adjust their relative placements. The recovered anchors provide the clean estimate for the next generation step (\textsection~\ref{sec:interfaces}). Figure~\ref{fig:overview} gives an overview.

\subsection{Local conformations and training targets}\label{sec:fit}\label{sec:target}

\begin{wrapfigure}[17]{r}{0.48\textwidth}
\centering
\includegraphics[width=\linewidth]{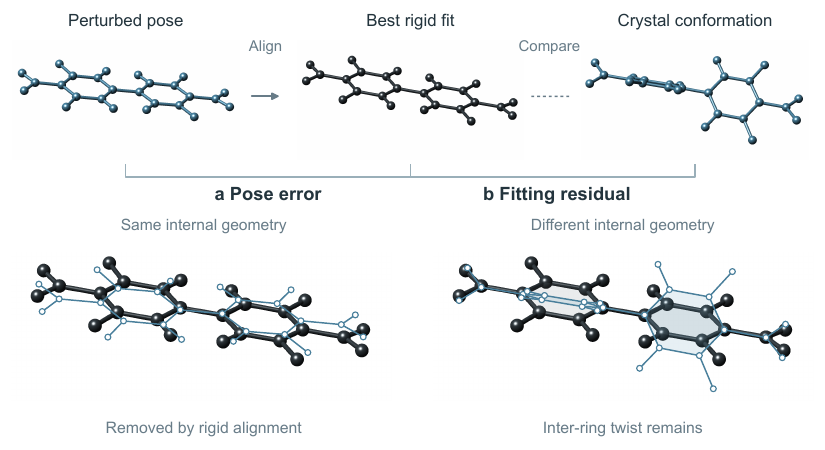}
\caption{\textbf{Pose error and conformation residual.} Rigid alignment removes pose error in \textbf{(a)}, but leaves the internal twist in \textbf{(b)}. Dark atoms show the rigid fit. Blue outlines show its perturbed or crystal counterpart.}
\label{fig:fitposeerrors}
\end{wrapfigure}
Moving and rotating a block cannot correct its internal conformation. We therefore use a separate fitting model to adjust RDKit conformations before assembly. It predicts torsions and applies equivariant coordinate corrections~\citep{landrum2006rdkit,kim2025mofflow2,satorras2021egnn}. We then freeze the model and its output geometry.

During training, we align all atoms of $\mathbf{X}_i$ with the corresponding crystal block to obtain $\mathbf{R}_i^*$. We define the clean anchor targets as $\mathbf{P}_i^*=\mathbf{R}_i^*\mathbf{A}_i$.

Direct crystal coordinates can have different inter-anchor distances because the fitted and crystal conformations retain internal differences. We therefore use the selected atoms under the best full-atom rigid fit as targets, keeping supervision consistent with rigid recovery. Because their coordinates are center-relative, we supervise translation separately through the center branch. Any remaining discrepancy between the rigidly aligned block and the crystal reflects internal conformation error that the assembly model cannot remove (Figure~\ref{fig:fitposeerrors}).

\finishwrap
\subsection{Anchor-state generation}\label{sec:anchors}

Building on periodic Bayesian flows for crystals~\citep{wu2025crysbfn}, we retain the Gaussian lattice branch and periodic center branch of MOF-BFN~\citep{jiao2025mofbfn} and replace its rotational state with a continuous Gaussian BFN over anchor coordinates~\citep{graves2023bfn}. During training, we sample the normalized anchor mean state at time $t\in[0,1]$ as
\begin{equation}
\boldsymbol{\mu}_t=\gamma_t\mathbf{P}^*/s+
\sqrt{\gamma_t(1-\gamma_t)}\,\boldsymbol{\epsilon},
\qquad
\gamma_t=1-\sigma_1^{2t},
\qquad
\boldsymbol{\epsilon}\sim\mathcal N(\mathbf{0},\mathbf{I}).
\label{eq:flow}
\end{equation}

We share the coordinate scale $s$ and terminal noise parameter $\sigma_1\in(0,1)$ across blocks. For $\gamma_t>0$, rescaling gives $s\boldsymbol{\mu}_t/\gamma_t=\mathbf{P}^*+\tau_t\boldsymbol{\epsilon}$, where $\tau_t^2=s^2(1-\gamma_t)/\gamma_t$.
Positive uniform scaling leaves the Kabsch orientation unchanged, so Proposition~1 describes the induced orientation uncertainty in the small-noise regime.

\begin{figure}[!t]
\centering
\includegraphics[width=\textwidth]{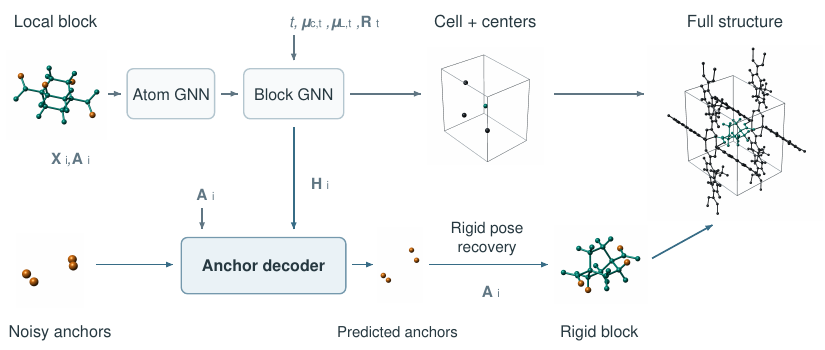}
\caption{\textbf{Anchor prediction and rigid recovery.} Local geometry, noisy anchors, and block context condition prediction. Alignment supplies $\mathbf{R}_t$ and recovers rigid blocks. Teal marks one block, black the others, and orange the anchors. Periodic images complete boundary-crossing blocks.}
\label{fig:architecture}
\end{figure}

To predict clean coordinates, the network combines each block's local geometry with the current crystal state (Figure~\ref{fig:architecture}). An Atom GNN first pools element and local-coordinate features into a block embedding. We also align $\mathbf{A}_i$ with $s\boldsymbol{\mu}_{i,t}$ to estimate the current rotation $\mathbf{R}_{i,t}$. A Block GNN takes a descriptor of this rotation, the block embedding, time, and center and lattice states as inputs and exchanges information across the structure. We pass its output $\mathbf{H}_i$ to the anchor decoder and use separate heads to predict the clean lattice and centers.

For each anchor, we construct a token from $\mathbf{H}_i$, its local coordinate $\mathbf{a}_{ik}$, its noisy coordinate $\boldsymbol{\mu}_{ik,t}$, and a periodic encoding of its current spatial position. The local coordinate identifies where the anchor lies within its block, while the noisy coordinate and periodic encoding describe its current placement. A Transformer exchanges information among anchors in the same structure. A shared pointwise head then predicts each anchor coordinate as
\begin{equation}
\widehat{\mathbf{p}}_{ik}^{\mathrm{raw}}
=s\left[\boldsymbol{\mu}_{ik,t}+g(\mathbf{z}'_{ik})\right],\qquad \mathbf{p}_{ik}^*=\mathbf{R}_i^*\mathbf{a}_{ik},
\label{eq:decode}
\end{equation}

Here $\mathbf{z}'_{ik}$ is the transformed feature and $g$ is the output head. The resulting coordinate prediction targets the clean endpoint $\mathbf{p}_{ik}^*$ at every step. We train with $\mathcal L_{\mathrm{pose}}=\mathcal L_{\mathrm{raw}}+\mathcal L_{\mathrm{rigid}}+\mathcal L_{\mathrm{proj}}$.
The coordinate terms supervise raw predictions and the rigid anchors obtained after recovery. The distance term penalizes deviations from local inter-anchor distances. Gradients pass through alignment and inter-block updates. Appendix~\ref{app:config} specifies the encodings, loss normalization, and weights.

\subsection{Inter-block coordination}\label{sec:interfaces}

We recover an initial rotation $\widehat{\mathbf{R}}_i^0$ by Kabsch alignment and obtain the Cartesian center $\widehat{\mathbf{c}}_i^0=\widehat{\mathbf{L}}\widehat{\mathbf{f}}_i$ from the center and lattice predictions. Placing each block with these estimates preserves its geometry, but does not enforce compatible placements between blocks. We therefore refine their relative poses using paired virtual points, following the paired-keypoint approach of EquiBind~\citep{stark2022equibind}.

Because inter-block connectivity is unknown, the network constructs these points from candidate block pairs using their current centers, local geometries, and learned embeddings. For each ordered pair $(i,j)$, we use four learned queries to attend to fixed local samples on block $i$ and predict virtual points $\mathbf{q}_{i\to j,m}$. We predict the corresponding points $\mathbf{q}_{j\to i,m}$ on block $j$ in the same way. The paired points represent locations that should coincide after assembly. They need not lie on atoms, and we predict a separate set for each block pair.

Given the current poses, the mismatch between paired points is
\begin{equation}
\mathbf{r}_{ijm}
=\mathbf{c}_j+\mathbf{L}\mathbf{n}_{ij}+\mathbf{R}_j\mathbf{q}_{j\to i,m}
-\mathbf{c}_i-\mathbf{R}_i\mathbf{q}_{i\to j,m},
\qquad
\mathbf{n}_{ij}=-\operatorname{round}\!\left(\mathbf{L}^{-1}(\mathbf{c}_j-\mathbf{c}_i)\right).
\label{eq:mainpairresidual}
\end{equation}

Here $\mathbf{L}$ is the predicted lattice, and $\mathbf{n}_{ij}$ indexes the image obtained by wrapping each fractional displacement component into the centered unit interval. The network predicts a confidence for each point pair. We weight the residuals by these confidences and aggregate them to compute translation increments. To compute rotation increments, we instead aggregate the confidence-weighted cross products of the center-relative virtual-point offsets with the residuals. After normalization and step limiting, we update the poses as
\begin{equation}
\mathbf{c}_i\leftarrow \mathbf{c}_i+\delta \mathbf{c}_i,
\qquad
\mathbf{R}_i\leftarrow\exp([\boldsymbol{\delta\omega}_i]_\times)\mathbf{R}_i,
\label{eq:mainposeupdate}
\end{equation}

where $[\boldsymbol{\delta\omega}_i]_\times$ is skew-symmetric. We apply two updates, recomputing residuals and periodic images while keeping local virtual points and confidences fixed. Losses on the corrected centers and rigid anchors train this step without connection labels (Appendix~\ref{app:config}).

During sampling, we use the recovered anchors $\widehat{\mathbf{P}}_i^{\mathrm{proj}}=\widehat{\mathbf{R}}_i\mathbf{A}_i$ as the clean estimate in Equation~\ref{eq:flow}, with the next noise level and a fresh Gaussian draw. Noise can distort the next state, so we recover rigid blocks at every step. The final lattice and corrected poses define the crystal, with atoms at $\widehat{\mathbf{y}}_{ia}=\widehat{\mathbf{c}}_i+\widehat{\mathbf{R}}_i\mathbf{x}_{ia}$.

\section{Experiments}
\label{sec:eval}

\subsection{Experimental setup}
\label{sec:protocol}
We evaluate on all 19,792 test structures of AtomMOF's MOF benchmark~\citep{kim2026atommof}, comparing with DiffCSP~\citep{jiao2023diffcsp} and MOF-specific block-level and all-atom methods.
For one or five candidates per structure, we report structure match rate (MR) under loose and strict criteria and normalized RMS displacement over successful matches, using \texttt{StructureMatcher}~\citep{ong2013pymatgen}.
Appendices~\ref{app:config} and~\ref{app:rms} specify the training settings and evaluation details.

\subsection{Structure reconstruction}
\begingroup
\setlength{\intextsep}{4pt}
\begin{table}[!t]
\centering\small
\setlength{\tabcolsep}{3.2pt}
\renewcommand{\arraystretch}{0.90}
\caption{\textbf{Generation granularity and structure recovery.} Block geom. denotes intra-block geometry. MR is in \%. Baseline results follow AtomMOF~\citep{kim2026atommof}.}
\label{tab:main}
\begin{tabular}{lcccccccc}
\toprule
&&&&&\multicolumn{2}{c}{Loose ($\stol=0.5$)}&\multicolumn{2}{c}{Strict ($\stol=0.3$)}\\
\cmidrule(lr){6-7}\cmidrule(lr){8-9}
Model&Block geom.&Granularity&Rigid&Samples&MR $\uparrow$&RMS $\downarrow$&MR $\uparrow$&RMS $\downarrow$\\
\midrule
\rowcolor{finegrain}
&&&&1&0.23&0.3896&0.01&0.1554\\
\rowcolor{finegrain}
\multirow[c]{-2}{*}{DiffCSP}&\multirow[c]{-2}{*}{No}&\multirow[c]{-2}{*}{Fine}&\multirow[c]{-2}{*}{No}&5&0.87&0.3982&0.08&0.1299\\
\rowcolor{coarsegrain}
&&&&1&21.93&0.3329&5.28&0.2036\\
\rowcolor{coarsegrain}
\multirow[c]{-2}{*}{MOFFlow}&\multirow[c]{-2}{*}{Yes}&\multirow[c]{-2}{*}{Coarse}&\multirow[c]{-2}{*}{Yes}&5&32.71&0.3290&8.68&0.2039\\
\rowcolor{coarsegrain}
&&&&1&27.66&0.3234&6.96&0.1979\\
\rowcolor{coarsegrain}
\multirow[c]{-2}{*}{MOF-BFN}&\multirow[c]{-2}{*}{Yes}&\multirow[c]{-2}{*}{Coarse}&\multirow[c]{-2}{*}{Yes}&5&40.15&0.3131&12.11&0.1950\\
\rowcolor{coarsegrain}
&&&&1&28.71&0.3094&8.20&0.1894\\
\rowcolor{coarsegrain}
\multirow[c]{-2}{*}{MOFFlow-2}&\multirow[c]{-2}{*}{Yes}&\multirow[c]{-2}{*}{Coarse}&\multirow[c]{-2}{*}{No$^\dagger$}&5&43.95&0.2925&15.98&0.1842\\
\rowcolor{atomgeometry}
&&&&1&28.63&0.2628&10.05&0.1328\\
\rowcolor{atomgeometry}
\multirow[c]{-2}{*}{AtomMOF-M}&\multirow[c]{-2}{*}{Yes}&\multirow[c]{-2}{*}{Fine}&\multirow[c]{-2}{*}{No}&5&46.08&0.2396&19.85&0.1249\\
\rowcolor{atomgeometry}
&&&&1&29.96&\textbf{0.2558}&11.07&\textbf{0.1275}\\
\rowcolor{atomgeometry}
\multirow[c]{-2}{*}{AtomMOF-L}&\multirow[c]{-2}{*}{Yes}&\multirow[c]{-2}{*}{Fine}&\multirow[c]{-2}{*}{No}&5&47.47&\textbf{0.2319}&21.17&\textbf{0.1199}\\
\rowcolor{mesograin}
&&&&1&\textbf{39.80}&0.2988&\textbf{12.36}&0.1794\\
\rowcolor{mesograin}
\multirow[c]{-2}{*}{\method}&\multirow[c]{-2}{*}{Yes}&\multirow[c]{-2}{*}{\textbf{Meso}}&\multirow[c]{-2}{*}{No}&5&\textbf{56.17}&0.2833&\textbf{21.20}&0.1774\\
\bottomrule
\end{tabular}
\par\vspace{3pt}
\begin{minipage}{\textwidth}\footnotesize
$^\dagger$MOFFlow-2 varies linker torsions while fixing bond lengths and angles. 
\end{minipage}
\end{table}
\endgroup

Table~\ref{tab:main} shows that finer generative resolution does not necessarily improve structure recovery.
By generating selected atomic positions and rigidly recovering complete blocks, AnchorPose combines spatially explicit pose modeling with constraints on internal geometry.
It surpasses the block-pose baselines in both match rates and improves single-candidate loose/strict MR over AtomMOF-L by $9.84/1.29$ percentage points.
With five candidates, the loose advantage remains $8.70$ points, while strict MR is essentially tied ($21.20$ versus $21.17$\%).
These gains show that a compact pose representation can recover more structures even when the assembled blocks cannot deform.
The advantage narrows as the matching tolerance tightens, however, and AtomMOF-L achieves lower RMS on successful matches.
This pattern suggests that the benefit of meso-grained generation lies primarily in reaching matchable assemblies, with further local refinement remaining a challenge for fixed conformations.
Because RMS is evaluated on different matched subsets, it cannot establish rigidity as the cause of the precision gap.
The results nevertheless support separating spatial pose generation from internal deformation, motivating the following analysis of how anchor states and rigid recovery contribute to reconstruction.

\subsection{What makes AnchorPose effective?}
\label{sec:mainablation}
We separate their role in the evolving state from their role in prediction, then examine how local fitting and inter-block coordination complement this representation (Table~\ref{tab:mainablation}).
\begingroup
\setlength{\intextsep}{4pt}
\begin{table}[H]
\centering\small
\caption{\textbf{Pose representation and component ablations.} All variants use actual block geometry. Without anchor states or predictions, models use Bingham states or rotation outputs, respectively. Checkmarks indicate enabled components. MR (\%) covers 19,792 structures with one candidate.}
\label{tab:mainablation}
\setlength{\tabcolsep}{4.2pt}
\begin{tabular}{lccccc rr}
\toprule
Variant & Anchor & Anchor & Geom. rot. & Block & Pose & Loose & Strict\\
& state & pred. & noise & fitting & updates & MR & MR\\
\midrule
\multicolumn{8}{l}{\emph{Anchor-state variants}}\\
\textbf{AnchorPose} & \checkmark & \checkmark & -- & \checkmark & \checkmark & \textbf{39.80} & \textbf{12.36}\\
Direct rotation prediction & \checkmark & & -- & \checkmark & \checkmark & 38.89 & 11.89\\
\midrule
\multicolumn{8}{l}{\emph{Rotation-state alternatives}}\\
Direct rotation prediction & & & & \checkmark & \checkmark & 36.41 & 9.86\\
Anchor prediction & & \checkmark & & \checkmark & \checkmark & 36.53 & 10.23\\
Geometry-aware rotation & & \checkmark & \checkmark & \checkmark & \checkmark & 35.05 & 10.00\\
\midrule
\multicolumn{8}{l}{\emph{Supporting-component ablations}}\\
Without block fitting & \checkmark & \checkmark & -- & & \checkmark & 34.00 & 9.82\\
Without pose updates & \checkmark & \checkmark & -- & \checkmark & & 37.81 & 11.11\\
\bottomrule
\end{tabular}
\end{table}
\endgroup

\paragraph{Anchor coordinates as the generative state.}
Predicting anchor coordinates is not sufficient on its own.
When the generative state remains a Bingham rotation, replacing direct rotation prediction with anchor prediction and Kabsch recovery changes loose/strict MR only from $36.41/9.86$ to $36.53/10.23$.
Using noisy anchor coordinates as the generative state raises MR to $38.89/11.89$ with direct rotation correction and $39.80/12.36$ with AnchorPose's anchor prediction and rigid recovery.
The gain therefore appears when anchor coordinates define the evolving pose state, rather than simply serving as the decoder output.

\paragraph{Geometry-aware rotation noise.}
One possible explanation for the advantage of AnchorPose is that the standard Bingham state uses geometry-independent rotational noise.
We test this alternative by deriving the Bingham noise from the full-block displacement tensor in Equation~\ref{eq:spatialmetric}, while retaining anchor prediction, fitted conformations, and inter-block pose updates.
This geometry-aware rotation state achieves $35.05/10.00$ loose/strict MR, compared with $36.53/10.23$ for the ordinary rotation state and $39.80/12.36$ for AnchorPose.
Explicitly incorporating block geometry into rotational noise therefore does not close the gap to an anchor-coordinate state.
Appendix~\ref{app:geometrybingham} provides the construction and calibration.

\paragraph{Complementary components.}
Local conformation fitting and inter-block pose updates address errors that the pose representation does not remove.
Using the original RDKit conformations lowers loose/strict MR from $39.80/12.36$ to $34.00/9.82$, showing that internal conformation quality is a major bottleneck for rigid assembly.
Removing the inter-block updates gives $37.81/11.11$, so coordinating neighboring blocks provides a smaller but consistent gain beyond independent Kabsch recovery.
All rotation-state comparisons retain both components, separating their effects from the contribution of the anchor-coordinate state.

\Needspace{10\baselineskip}
\subsection{Anchor Count}
\label{sec:components}
\begingroup
\setlength{\intextsep}{3pt}
\begin{table}[H]
\begin{minipage}[t]{0.46\textwidth}
\vspace{0pt}
We vary the anchor budget from four to eight (Table~\ref{tab:anchorcount}).
Four anchors give the highest loose and strict match rates, with both decreasing slightly at larger budgets.
The additional point predictions thus provide no accuracy benefit in this comparison.
We retain four anchors as a compact pose state.
\end{minipage}\hfill
\begin{minipage}[t]{0.49\textwidth}
\vspace{0pt}
\centering\small
\setlength{\abovecaptionskip}{0pt}
\setlength{\belowcaptionskip}{4pt}
\caption{\textbf{Single-candidate MR (\%).}}
\label{tab:anchorcount}
\begin{tabular*}{\linewidth}{@{\extracolsep{\fill}}lrrr@{}}
\toprule
Anchors & 4 & 6 & 8\\
\midrule
Loose MR & \textbf{39.80} & 39.39 & 38.82\\
Strict MR & \textbf{12.36} & 12.31 & 12.00\\
\bottomrule
\end{tabular*}
\end{minipage}
\end{table}
\endgroup

\subsection{Pose-State Analysis}
\label{sec:main_diagnostics}
Figure~\ref{fig:diagnostics} reveals a qualitative difference between rotation and anchor-coordinate states.
A Bingham state produces similar angular uncertainty across block sizes.
Because the same angular deviation moves atoms farther in a larger block, its spatial perturbation grows substantially with block size.
Under anchor-coordinate noise, larger blocks exhibit smaller angular deviations, while atomic displacements remain comparable across sizes.
With anchor states, the denoiser therefore receives approximately equal spatial pose errors across block sizes.

Consistent with Proposition~1, spatially extended anchors constrain orientation more tightly under the same positional noise.
The approximately balanced full-atom errors are an empirical observation, since shapes and anchor coverage also vary across groups.
These observations support expressing pose uncertainty in the Cartesian coordinates used to evaluate reconstruction.
Under the native noise schedules, Bingham has lower spatial errors at earlier time points, while Gaussian anchors have lower input and recovered-pose errors at later time points.
\begingroup
\setlength{\intextsep}{4pt}
\begin{figure}[H]
\centering
\includegraphics[width=0.94\textwidth]{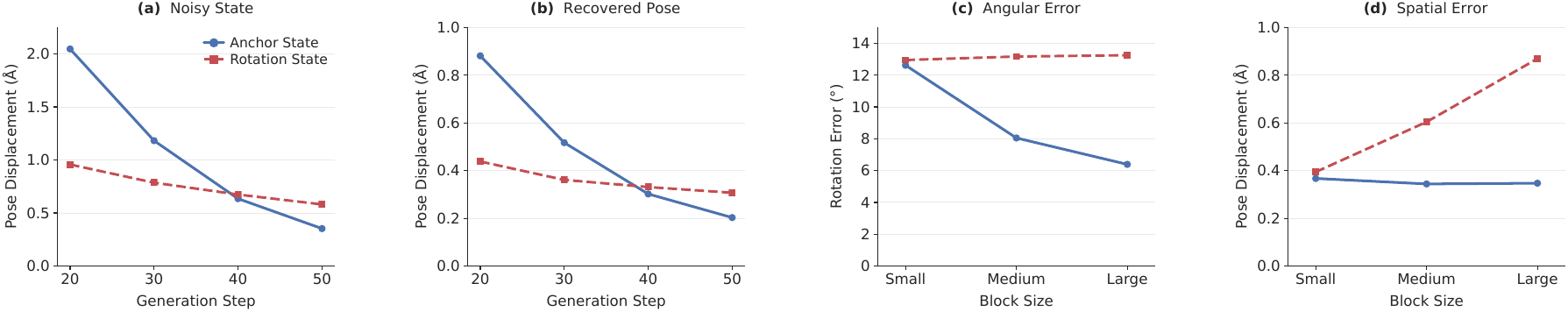}
\caption{\textbf{Pose errors under the native Gaussian-anchor and Bingham schedules.} Medians over the paired validation probe. Panels (a,b) show input and recovered-pose errors over time. Panels (c,d) show step-50 input errors by block size. Spatial errors measure full-template RMS displacement from the target rigid pose. Size groups are tertiles of the full-atom RMS radius. Recovered poses include Kabsch alignment and inter-block updates. Appendix~\ref{app:posecontrols} details the probe protocol.}
\label{fig:diagnostics}
\label{fig:stategeometry}
\end{figure}
\endgroup

\section{Discussion and conclusion}
AnchorPose treats MOF assembly as geometry-aware pose generation at a meso-grained level.
The geometry dependence of rotation errors motivates generating positions of representative atoms.
These corresponding points connect the current state to spatial prediction and supervision, while rigid recovery preserves complete building blocks and supplies the next step's clean estimate.
Together, these choices incorporate block geometry throughout pose generation and improve structure match rates while generating only a few points per block.

AnchorPose holds fitted block conformations fixed during assembly, so pose generation cannot correct their remaining internal errors. Nearly collinear anchors also provide weak constraints on rotation about the anchor line.
Because each anchor corresponds to a specific atom, symmetry-equivalent poses can still have different coordinate representations.
Future work could use directional coverage to guide anchor selection and account for equivalent poses during training. Jointly modeling conformation and assembly would also allow internal geometry to adapt during block placement.

\section*{Reproducibility statement}
Appendix~\ref{app:theory} derives the geometric identities and states their assumptions.
Appendix~\ref{app:config} specifies the architecture, rigid recovery, losses, and training and sampling procedures, together with the evaluation protocol.
Appendix~\ref{app:posecontrols} records the supplementary results and controlled configurations. Appendix~\ref{app:examples} presents qualitative examples.
The full implementation, including training and evaluation code, will be publicly released upon acceptance.

\section*{AI use statement}
We used AI assistants for literature discovery, methodological discussion and derivation checks, implementation and experiment analysis, manuscript drafting, and figure-generation code.
The reported numerical results originate from model runs and recorded evaluations, not generated measurements.
The authors have checked the reported results and take responsibility for their accuracy.

\begingroup
\setlength{\bibsep}{2pt}
\bibliography{references}
\bibliographystyle{iclr2027_conference}
\endgroup
\clearpage
\appendix
\section{Geometric derivations}
\label{app:theory}

\subsection{From rotation error to atomic displacement}
For a centered template $\mathbf{X}=\{\mathbf{x}_a\}_{a=1}^n$, the squared displacement at a fixed block center is
\begin{equation}
 d_{\mathbf{X}}^2(\mathbf{R},\widehat{\mathbf{R}})=\frac1n\sum_a\|\widehat{\mathbf{R}} \mathbf{x}_a-\mathbf{R}\mathbf{x}_a\|^2.
 \label{eq:spatialmetric_app}
\end{equation}
Let $\mathbf{Q}=\mathbf{R}^\top\widehat{\mathbf{R}}$ rotate through angle $\theta$ about unit axis $\mathbf{u}$.
The component of $\mathbf{x}_a$ parallel to $\mathbf{u}$ stays unchanged, while its perpendicular component traces a chord. Therefore,
\begin{align}
 \|(\mathbf{Q}-\mathbf{I})\mathbf{x}_a\|^2&=4\sin^2(\theta/2)\,\mathbf{u}^\top(\|\mathbf{x}_a\|^2\mathbf{I}-\mathbf{x}_a\mathbf{x}_a^\top)\mathbf{u},\\
 d_{\mathbf{X}}^2(\mathbf{R},\widehat{\mathbf{R}})&=4\sin^2(\theta/2)\,\mathbf{u}^\top \mathbf{J}_{\mathbf{X}}\mathbf{u},
 \qquad \mathbf{J}_{\mathbf{X}}=\frac1n\sum_a(\|\mathbf{x}_a\|^2\mathbf{I}-\mathbf{x}_a\mathbf{x}_a^\top).
\end{align}
This proves Equation~\ref{eq:spatialmetric} for arbitrary angles and agrees with shape-dependent rigid-pose distances~\citep{popov2014rigidrmsd,bregier2017pose}.
For the body-frame perturbation $\widehat{\mathbf{R}}=\mathbf{R}\exp([\boldsymbol{\delta\omega}]_\times)$,
\begin{equation}
 d_{\mathbf{X}}^2(\mathbf{R},\widehat{\mathbf{R}})=\boldsymbol{\delta\omega}^\top \mathbf{J}_{\mathbf{X}}\boldsymbol{\delta\omega}+O(\|\boldsymbol{\delta\omega}\|^3).
 \label{eq:metric_app}
\end{equation}
Scaling the block by $r$ scales $\mathbf{J}_{\mathbf{X}}$ by $r^2$. Its eigenvalues describe sensitivity about different axes.
Thus larger spatial extent requires finer angular accuracy for the same Cartesian tolerance.

\subsection{Orientation uncertainty induced by noisy points}
\label{app:pointuncertainty}
\paragraph{Proof of Proposition 1.}
Suppose we observe corresponding anchors as $\mathbf{z}_k=\mathbf{R}^*\mathbf{a}_k+\tau\boldsymbol{\epsilon}_k$, with independent $\boldsymbol{\epsilon}_k\sim\mathcal N(\mathbf{0},\mathbf{I})$.
Kabsch centers both point sets. Write $\bar{\mathbf{a}}_k=\mathbf{a}_k-K^{-1}\sum_l \mathbf{a}_l$ and $\mathbf{e}_k=(\mathbf{R}^*)^\top(\boldsymbol{\epsilon}_k-K^{-1}\sum_l\boldsymbol{\epsilon}_l)$.
Linearization around $\mathbf{R}^*$ yields
\begin{equation}
 \min_{\boldsymbol{\delta\omega}}\sum_k\|\tau \mathbf{e}_k-[\boldsymbol{\delta\omega}]_\times\bar{\mathbf{a}}_k\|^2,
 \qquad \mathbf{J}_{\mathbf{A}}\boldsymbol{\delta\omega}=\tau\sum_k\bar{\mathbf{a}}_k\times \mathbf{e}_k,
\end{equation}
where
\begin{equation}
 \mathbf{J}_{\mathbf{A}}=\sum_k(\|\bar{\mathbf{a}}_k\|^2\mathbf{I}-\bar{\mathbf{a}}_k\bar{\mathbf{a}}_k^\top),
 \qquad \operatorname{Cov}(\boldsymbol{\delta\omega})\simeq\tau^2\mathbf{J}_{\mathbf{A}}^{-1}.
 \label{eq:uncertainty_app}
\end{equation}
The covariance follows because $\sum_k\bar{\mathbf{a}}_k=\mathbf{0}$ and the right-hand side has covariance $\tau^2\mathbf{J}_{\mathbf{A}}$.
The approximation assumes small noise and non-collinear anchors. Collinearity leaves rotation about the anchor line unidentifiable, and near-collinearity makes the estimate poorly conditioned.

$\mathbf{J}_{\mathbf{X}}$ averages all atoms about the full-block centroid, whereas $\mathbf{J}_{\mathbf{A}}$ sums selected anchors about their own centroid.
The former governs full-atom spatial error, while the latter governs local orientation uncertainty induced by anchor noise.
Since $\mathbf{u}^\top \mathbf{J}_{\mathbf{A}}\mathbf{u}=\sum_k\|\mathbf{u}\times\bar{\mathbf{a}}_k\|^2$, long perpendicular lever arms constrain rotation more tightly under a given level of positional noise.
The anchor layout therefore determines which orientation changes positional noise can obscure.

\subsection{Connection to the Gaussian anchor state}
\label{app:anchorstate}
For the implemented flow in Equation~\ref{eq:flow}, at $\gamma_t>0$,
\begin{equation}
 \mathbf{Z}_t=\frac{s}{\gamma_t}\boldsymbol{\mu}_t=\mathbf{P}^*+\tau_t\boldsymbol{\epsilon},
 \qquad \tau_t^2=s^2\frac{1-\gamma_t}{\gamma_t}.
\end{equation}
Positive uniform scaling does not change the ideal Kabsch orientation, so Equation~\ref{eq:uncertainty_app} describes its local uncertainty.
The network receives the original $\boldsymbol{\mu}_t$, including its shrinkage. At $\gamma_t=0$ there is no orientation observation, and the local approximation does not apply.
The standard Bingham controls use a common geometry-independent rotation schedule with terminal parameter $\beta=200$ and 50 steps.
Rendering their rotation as $\mathbf{R}_t\mathbf{A}$ exposes block geometry to the decoder without altering that schedule.
Appendix~\ref{app:matchedprotocol} also evaluates a geometry-dependent Bingham state.
These relations concern input uncertainty. Appendix~\ref{app:smallblocks} reports prediction and sampling measurements separately.

\begin{figure}[!htbp]
\centering
\includegraphics[width=\textwidth]{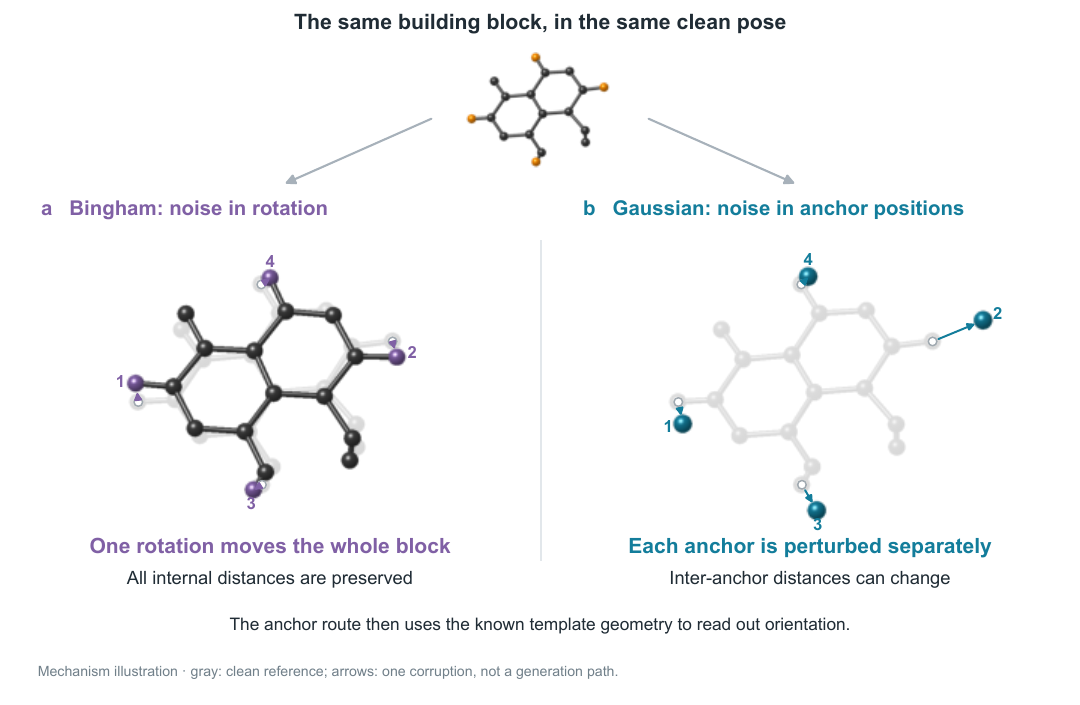}
\caption{\textbf{Noise acts on different pose variables.} From the same clean block, \textbf{(a)} a noisy Bingham orientation moves the template rigidly, whereas \textbf{(b)} Gaussian anchors move independently. Gray denotes the clean reference. Panel (b) undoes scale and time-dependent shrinkage without rigid projection. Arrows show one native corruption draw.}
\label{fig:posenoisemechanism}
\end{figure}

Figure~\ref{fig:posenoisemechanism} illustrates the distinction between native rotation and point noise.

\subsection{The selected-point metric and rigid recovery}
Equation~\ref{eq:pointmetric} is the Euclidean distance between two valid rigid representations with known atom identities.
Like Equation~\ref{eq:spatialmetric}, it weights angular error according to the represented points, but uses only the selected atoms.
Noisy states and raw predictions can lie outside the set $\{\mathbf{R}\mathbf{A}\mid \mathbf{R}\in\SO(3)\}$.
Centered Kabsch finds a best-fit rotation after removing each point set's centroid. Reconstructing $\widehat{\mathbf{R}} \mathbf{A}$ restores the anchor layout about the original block center.
The separate center head determines translation.
Thus raw-coordinate supervision, centered rigid alignment, and the metric between reconstructed poses have different roles.

\subsection{Anchor selection and correspondence}
\label{app:representation}
FPS starts with the atom farthest from the full-block centroid, then repeatedly selects the atom farthest from the selected set. We retain up to four atoms, masking padding for smaller blocks.
Four point coordinates provide a redundant, twelve-dimensional spatial representation of block orientation.
Alignment maps the predictions to $\SO(3)$, and reconstruction preserves the complete block's rigid geometry.
Three non-collinear corresponding points identify a proper rotation in the noiseless case.
We use four FPS atoms to obtain a small, spatially dispersed set. Their layout determines the orientation constraints through $\mathbf{J}_{\mathbf{A}}$.

For rotation through angle $\theta$ about an axis, a point at perpendicular distance $r_\perp$ moves by $2r_\perp|\sin(\theta/2)|$.
At small angles, the same positional tolerance $\varepsilon$ therefore imposes an angular tolerance of approximately $\varepsilon/r_\perp$.
Long lever arms make rotation error easier to detect in point coordinates and constrain it more tightly during alignment.
Spatial dispersion supplies these constraints in several directions. Distance from the block center alone does not determine distance from every rotation axis.
Once corresponding non-collinear anchors determine a rotation, applying it to the fixed template places all remaining atoms with that same rigid transformation.
Our four-point choice concentrates the generated coordinates on this spatial coverage rather than on densely sampling the block interior.

The derivations and training targets use fixed atom identities.
Symmetry-aware pose distances can additionally minimize over equivalent proper transformations~\citep{bregier2017pose}.
The benchmark instead performs periodic structure matching as specified in Appendix~\ref{app:rms}.

\paragraph{Atom-order check.}
FPS selects atom identities rather than assigning new correspondences. The selected coordinates, noisy states, and targets retain the same identities.
Equal-distance ties are resolved by input order.
We randomly permuted the atoms once per block occurrence across the full test set and repeated the implemented FPS selection.
The unordered selected-atom set changed in 27 of 123,861 occurrences (0.022\%).
For a fixed selected set, the anchor decoder uses shared token mappings without index embeddings, so jointly permuting the tokens and their associated inputs permutes the point outputs.

\subsection{Full-test anchor geometry statistics}
\label{app:anchorcoverage}
We measure geometric coverage on all 19,792 test structures, comprising 123,861 block occurrences, without filtering by prediction success.
Of these, 85,419 have non-collinear full templates. FPS preserves non-collinearity in every such case.
The remaining templates are single-point or collinear, for which some rotations leave atom positions unchanged.

To compare directional sensitivity rather than overall scale, define $\mathbf{M}_{\mathbf{X}}=\mathbf{J}_{\mathbf{X}}/\operatorname{tr}(\mathbf{J}_{\mathbf{X}})$ and $\mathbf{M}_{\mathbf{A}}=\mathbf{J}_{\mathbf{A}}/\operatorname{tr}(\mathbf{J}_{\mathbf{A}})$, using the separately centered templates and anchors above.
For non-collinear templates, let
\begin{equation}
 \mathbf{B}=\mathbf{M}_{\mathbf{X}}^{-1/2}\mathbf{M}_{\mathbf{A}}\mathbf{M}_{\mathbf{X}}^{-1/2},\qquad
 D=\frac{\lambda_{\max}(\mathbf{B})}{\lambda_{\min}(\mathbf{B})}.
\end{equation}
The eigenvalues are the extrema of $\mathbf{u}^\top \mathbf{M}_{\mathbf{A}}\mathbf{u}/(\mathbf{u}^\top \mathbf{M}_{\mathbf{X}}\mathbf{u})$ over rotation axes.
Thus $D=1$ means that anchors reproduce the full template's relative directional sensitivity exactly. Larger values indicate greater distortion.
Table~\ref{tab:anchorcoverage} evaluates the 85,375 non-collinear block occurrences with four anchors, comparing FPS with one uniformly sampled four-atom subset per occurrence.
FPS better preserves directional sensitivity in the median and upper quantiles, although a tail of poorly matched geometries remains.

\begin{table}[htbp]
\centering\small
\caption{\textbf{Directional geometric coverage of four anchors.} Distortion $D$ is scale-normalized. Lower is better. Statistics count block occurrences across the full test set.}
\label{tab:anchorcoverage}
\begin{tabular}{lrrr}
\toprule
Selection & Median & 90th percentile & 95th percentile\\
\midrule
FPS & 1.66 & 3.06 & 10.34\\
Random four atoms & 3.80 & 16.59 & 29.01\\
\bottomrule
\end{tabular}
\end{table}

\subsection{Geometry-dependent Bingham control}
\label{app:geometrybingham}
The rotation-state control in Table~\ref{tab:mainablation} uses the full fitted block to define a Bingham sender distribution.
For the centered coordinates $\mathbf{X}_i$, let $\mathbf{G}_i=\mathbf{J}_{\mathbf{X}_i}/c$, where $\mathbf{J}_{\mathbf{X}_i}$ is the all-atom tensor in Equation~\ref{eq:spatialmetric}.
We fix $c$ to the mean of $\operatorname{tr}(\mathbf{J}_{\mathbf{X}})/3$ over training-block occurrences, giving $c=6.555835183555313~\text{\AA}^2$.
This common normalization preserves differences in block size and directional sensitivity.

Write unit quaternions in $(x,y,z,w)$ order, and let $\mathbf{v}(\mathbf{q}^{-1}\otimes \mathbf{y})$ denote the vector part of the relative quaternion.
For clean orientation $\mathbf{q}$ and noisy observation $\mathbf{y}$, sender step $j$ uses
\begin{equation}
 p_{i,j}(\mathbf{y}\mid \mathbf{q})\propto
 \exp\!\left[-\lambda\alpha_j \mathbf{v}(\mathbf{q}^{-1}\otimes \mathbf{y})^\top \mathbf{G}_i \mathbf{v}(\mathbf{q}^{-1}\otimes \mathbf{y})\right]
 =\exp\!\left[-\frac{\lambda\alpha_j}{4c}d_{\mathbf{X}_i}^2(\mathbf{R}(\mathbf{q}),\mathbf{R}(\mathbf{y}))\right].
 \label{eq:geometrybingham_sender}
\end{equation}
Here $\lambda$ is a global precision multiplier and $\alpha_j$ is the base sender precision.
The equality follows from Equation~\ref{eq:spatialmetric}, since the relative quaternion has vector part $\sin(\theta/2)\mathbf{u}$.
Thus the sender penalizes the actual mean squared displacement of all block atoms, including their different distances from each rotation axis.

Let $\boldsymbol{\mathcal L}(\mathbf{q})$ be the orthogonal matrix satisfying $\boldsymbol{\mathcal L}(\mathbf{q})\mathbf{y}=\mathbf{q}\otimes \mathbf{y}$, and define
\begin{equation}
 \mathbf{H}_i(\mathbf{q})=-\boldsymbol{\mathcal L}(\mathbf{q})\operatorname{diag}(\mathbf{G}_i,0)\boldsymbol{\mathcal L}(\mathbf{q})^\top,
 \qquad
 p_{i,j}(\mathbf{y}\mid \mathbf{q})\propto\exp\!\left[\lambda\alpha_j \mathbf{y}^\top \mathbf{H}_i(\mathbf{q})\mathbf{y}\right].
 \label{eq:geometrybingham_matrix}
\end{equation}
This gives the Bingham natural-parameter matrix directly from block geometry.
Its normalizer is independent of $\mathbf{q}$, and its likelihood is symmetric in $\mathbf{q}$ and $\mathbf{y}$.
For independent observations $\mathbf{y}_1,\ldots,\mathbf{y}_{n-1}$ and a uniform prior, the state before decoder call $n$ is
\begin{equation}
 p_{i,n}(\mathbf{q})\propto\exp(\mathbf{q}^\top \mathbf{C}_{i,n}\mathbf{q}),\qquad
 \mathbf{C}_{i,n}=\sum_{j=1}^{n-1}\lambda\alpha_j\mathbf{H}_i(\mathbf{y}_j),\qquad \mathbf{C}_{i,1}=\mathbf{0}.
 \label{eq:geometrybingham_posterior}
\end{equation}
The eigenvector of $\mathbf{C}_{i,n}$ with largest eigenvalue supplies the current orientation estimate.
For $\mathbf{G}_i=\mathbf{I}$, $\mathbf{H}_i(\mathbf{q})=\mathbf{q}\mathbf{q}^\top-\mathbf{I}_4$, which recovers the ordinary Bingham sender up to an irrelevant scalar identity term.
Degenerate templates retain their zero geometric eigenvalues and associated unobservable rotations.

The control uses $N=50$ decoder calls and constant base sender precision $\alpha_j=4$, with $\lambda=1.1387886347566916$ at every step.
Call $n$ therefore uses $n-1$ observations. The final input state uses 49.
Training simulates this state around the target quaternion.
During generation, the endpoint-feedback sampler simulates the next state around the latest recovered clean quaternion, retaining the same $\mathbf{G}_i$ and precision schedule.
The model retains the point decoder, coordinate objectives, and inter-block updates described in Appendix~\ref{app:matchedprotocol}.

We calibrate $\lambda$ on 256 evenly spaced training structures containing 1,696 blocks, with four noise draws per block.
The target is the ordinary Bingham process's block-averaged full-atom spatial MSE at decoder call 50.
Nine geometric-bisection trials over $[0.125,8]$ select the multiplier with the smallest relative discrepancy.
The resulting MSEs are $0.461886$ and $0.459698~\text{\AA}^2$ for geometry-dependent and ordinary Bingham, respectively, a 0.48\% difference.
This calibration matches one terminal spatial-noise statistic against ordinary Bingham. It does not match the full trajectory or the Gaussian-anchor process.
We fix the calibrated multiplier before training, without validation- or test-match-rate tuning.

\FloatBarrier
\section{Implementation and evaluation}
\label{app:config}

\subsection{Network and training settings}
We train the fitting network first and freeze its parameters, centered outputs $\mathbf{X}_i$, FPS anchors $\mathbf{A}_i$, and sixteen-point shape samples throughout assembly training and generation.
Each fitted template retains its preprocessing coordinate frame and atom correspondence. Assembly recenters it without assigning a new canonical orientation.
Table~\ref{tab:settings} summarizes the complete model. Appendix~\ref{app:posecontrols} specifies the controlled variants.
\begin{table}[htbp]
\centering\small
\caption{\textbf{Training and architecture settings.} Batches are effective global batches. Both stages use BF16 mixed precision and AdamW.}
\label{tab:settings}
\begin{tabular}{p{0.29\textwidth}p{0.64\textwidth}}
\toprule
Setting&Value\\
\midrule
Local-fitting network&Width 384, four atom-message layers, simultaneous torsion head, four coordinate-residual layers\\
Local-fitting loss&Circular torsion loss and aligned linker squared error, each with weight 1\\
Local-fitting optimization&Learning rate $10^{-4}$, cosine decay to $10^{-6}$, weight decay $10^{-5}$, gradient clipping 5\\
Local-fitting budget&200 epochs, 260 updates/epoch, batch 768, seed 348, validation-selected checkpoint at epoch index 199\\
\midrule
Assembly atom / block GNN&Four atom layers, width 64, six block layers, width 512, time embedding width 128\\
Anchor decoder&Up to four FPS atoms, four Transformer layers, width 128, coordinate scale $s=10$~\AA\\
Rigid recovery&Kabsch alignment, two inter-block point-based updates (Appendix~\ref{app:solver})\\
BFN schedules&50 steps, anchor $\sigma_1=0.05$, lattice $\sigma_1=0.0316228$, periodic-center precision parameter 1,000\\
Assembly loss weights&Lattice 1, center 1, pose 4, raw, rigidity, and projected pose terms each have internal weight 1\\
Assembly optimization&Learning rate $5\times10^{-4}$, betas $(0.9,0.98)$, weight decay 0.01, gradient clipping 0.5\\
Assembly schedule&Reduce-on-plateau with factor 0.6, patience 30, floor $10^{-4}$\\
Assembly budget&1,000 epochs, 50 updates/epoch, batch 4,032, seed 123, EMA decay 0.995 from step zero\\
Evaluation checkpoint&Final EMA at epoch index 999, update 50,000, 50 generation steps, seed 123 for one candidate\\
\bottomrule
\end{tabular}
\end{table}

\subsection{Model size and inference cost}
\label{app:cost}
Table~\ref{tab:cost} reports measurements on one NVIDIA A100 80GB PCIe GPU with PyTorch 2.0.1 and batch size 32.
We use 256 validation structures selected at evenly spaced indices, repeat each batch three times, and exclude two warm-up batches per stage.
CUDA synchronization brackets each timed call. Times cover the fitting-network forward pass or all 50 assembly steps and final atomic-coordinate reconstruction, respectively. Data loading, RDKit conformer generation, file writing, and structure matching are excluded.
Fitting uses BF16 autocast. Assembly uses FP32 with TF32 matrix multiplication enabled.
The two stages are profiled separately, and memory is the peak allocated GPU memory, including model parameters and inputs.
Fitted conformations are reused across assembly candidates.

\begin{table}[htbp]
\centering\small
\caption{\textbf{Model size and inference cost.} Mean time per batch of 32 structures. Peak allocated memory over the profiled batches.}
\label{tab:cost}
\begin{tabular}{lrrr}
\toprule
Stage & Parameters (M) & Time / batch (s) & Peak memory (GiB)\\
\midrule
Local fitting & 8.34 & 0.059 & 0.92\\
Assembly (50 steps) & 11.99 & 1.181 & 0.28\\
\bottomrule
\end{tabular}
\end{table}

\subsection{Local fitting and rigid targets}
\label{app:fitdetails}
The fitting Atom GNN encodes elements, formal charges, chirality, bond types, and intra-block distances.
Pooled block features and a composition summary condition a shared torsion head.
For each ordered four-atom tuple, forward and reversed atom features predict a two-dimensional residual given by
\begin{equation}
 \mathbf{u}_k^0=(\cos\phi_k^0,\sin\phi_k^0),\qquad
 \widehat{\mathbf{u}}_k=\operatorname{normalize}(\mathbf{u}_k^0+\Delta \mathbf{u}_k),\qquad
 \widehat\phi_k=\operatorname{atan2}(\widehat u_{k,2},\widehat u_{k,1}).
\end{equation}
The head predicts all angles together. Preprocessing selects a graph-center root by minimum maximum bond-hop distance, and identifies the rotating subtree after removing each modeled bond.
We apply wrapped differences $\operatorname{wrap}(\widehat\phi_k-\phi_k^0)$ in increasing subtree depth, using the updated bond axes.

Four equivariant coordinate-residual layers then refine linker atoms using intra-block bonds and radius edges~\citep{satorras2021egnn}.
We mask metal nodes. Linkers without modeled torsions can still receive coordinate updates.
A rigid alignment removes global drift introduced by this residual branch.
The loss combines circular torsion error with the mean linker error
\begin{equation}
 D_i(\widetilde{\mathbf{X}}_i,\mathbf{Y}_i)=\min_{\mathbf{R}\in\SO(3),\mathbf{c}\in\R^3}
 \frac1{n_i}\sum_a\|\mathbf{R}\widetilde{\mathbf{x}}_{ia}+\mathbf{c}-\mathbf{y}_{ia}\|^2,
 \label{eq:fiterror}
\end{equation}
where $\widetilde{\mathbf{X}}_i$ is the fitting output before centering.
The aligned loss backpropagates through both branches. We omit empty torsion terms.
Crystal blocks retain their supplied Cartesian image assignments before alignment, without independently wrapping their atoms across cell faces.

\begin{figure}[htbp]
\centering
\includegraphics[width=0.94\textwidth]{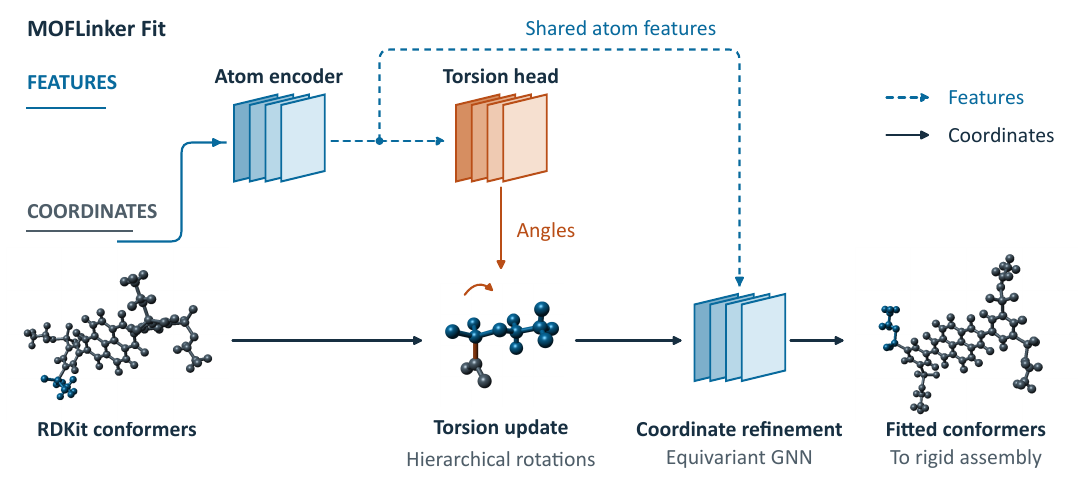}
\caption{\textbf{Local conformation preparation.} Shared atom features predict torsions, followed by equivariant coordinate refinement. Assembly uses the resulting fixed block.}
\label{fig:fit_architecture}
\end{figure}

For the centered fitted block $\mathbf{X}_i$ and centered crystal coordinates $\bar{\mathbf{y}}_{ia}^*$, full-atom alignment defines the assembly targets as
\begin{equation}
 \mathbf{R}_i^*=\arg\min_{\mathbf{R}\in\SO(3)}\sum_a\|\mathbf{R}\mathbf{x}_{ia}-\bar{\mathbf{y}}_{ia}^*\|^2,
 \qquad \mathbf{P}_i^*=\mathbf{R}_i^*\mathbf{A}_i.
 \label{eq:target}
\end{equation}
We distinguish local fitting from pose prediction using two complementary errors
\begin{equation}
 e_{i,\mathrm{conf}}^2=\frac1{n_i}\sum_a\|\mathbf{R}_i^*\mathbf{x}_{ia}-\bar{\mathbf{y}}_{ia}^*\|^2,\qquad
 e_{i,\mathrm{pose}}^2=\frac1{n_i}\sum_a\|\widehat{\mathbf{R}}_i\mathbf{x}_{ia}-\mathbf{R}_i^*\mathbf{x}_{ia}\|^2.
 \label{eq:conformationposeerrors}
\end{equation}
For fixed atom correspondences, $e_{i,\mathrm{conf}}$ is the lowest RMSD achievable by rigidly placing that fitted conformation.
Local fitting reduces this error floor. Assembly controls $e_{i,\mathrm{pose}}$, measured at a common center.
Their squares are not additive components of total error, because the total coordinate residual also contains a cross term.

Figure~\ref{fig:fitposeerrors} in Section~\ref{sec:fit} illustrates this distinction.

\subsection{Block features and anchor tokens}
The assembly Atom GNN pools local atom features into block embeddings.
The Block GNN combines these with time, center and lattice states, and the current orientation descriptor.
For Gaussian anchors, noisy-point alignment gives a quaternion $\mathbf{q}_i$.
Let $r(\mathbf{B})=\sqrt{K_i^{-1}\sum_k\|\mathbf{b}_k-\bar{\mathbf{b}}\|^2}$ denote a point set's centered RMS radius.
The descriptor is
\begin{equation}
 \mathbf{h}_i^{pose}=\alpha_i\operatorname{vec}(\mathbf{q}_i\mathbf{q}_i^\top),\qquad
 \alpha_i=o_i\operatorname{clip}_{[0,1]}
 \left(\frac{r(s\boldsymbol{\mu}_{i,t})}{\max(r(\mathbf{A}_i),10^{-6}\,\text{\AA})}\right),
\end{equation}
where $o_i$ is one for an observable template and zero otherwise.
This strength scales the orientation feature using the extent of the current point state.

Each anchor receives
\begin{align}
 \mathbf{z}_{ik}&=\mathbf{W}_H\mathbf{H}_i+E_A(\mathbf{a}_{ik}/s)+E_\mu(\boldsymbol{\mu}_{ik,t})+E_\psi(\boldsymbol{\psi}_{ik})+\mathbf{e}_{ik},\label{eq:token}\\
 \boldsymbol{\psi}_{ik}&=[\sin(2\pi\boldsymbol{\xi}_{ik}),\cos(2\pi\boldsymbol{\xi}_{ik})],\qquad
 \boldsymbol{\xi}_{ik}=\widehat{\mathbf{f}}_i+\widehat{\mathbf{L}}^{-1}(s\boldsymbol{\mu}_{ik,t}).\label{eq:phase}
\end{align}
The projections have width 128. $\mathbf{e}_{ik}$ encodes the anchor-kind flag.
Anchor tokens have no index embeddings. Their coordinates, states, and targets share the same selected atom identities. FPS resolves equal-distance ties by input order.
Four Transformer layers allow anchors within the same MOF to interact, and a shared MLP predicts the clean-point residual in Equation~\ref{eq:decode}.
Local coordinates stay continuous when a block crosses a cell face. The global-position encoding is periodic under integer lattice shifts.

\paragraph{Alignment and degenerate blocks.}
We detach noisy-state alignment and differentiate through output alignment.
We use float32 SVD on the centered cross-covariance $\mathbf{C}$ with regularizer
$10^{-7}\max(\|\mathbf{C}\|_F,1)\operatorname{diag}(1,2,3)$ and a determinant correction to obtain a proper rotation.
A template is observable if its second centered singular value exceeds $10^{-6}\max(\sigma_{\max},10^{-8})$.
Otherwise its input pose descriptor has zero strength.
The output decoder retains the regularized SVD result. It selects identity for a one-point centered template, while rotation about a collinear template's line remains underdetermined.
Attention and alignment exclude padding.

\subsection{BFN states and training objectives}
Equation~\ref{eq:flow} defines the anchor state conditional on clean center-relative coordinates $\mathbf{P}^*$.
The assembly loss is
\begin{align}
 \mathcal L&=\mathcal L_{\mathrm{lattice}}+\mathcal L_{\mathrm{center}}+4\mathcal L_{\mathrm{pose}},\label{eq:loss}\\
 \mathcal L_{\mathrm{pose}}&=\mathcal D_t(\widehat{\mathbf{P}}^{raw}/s,\mathbf{P}^*/s)
 +\mathcal L_{\mathrm{rigid}}
 +\mathcal D_t(\widehat{\mathbf{P}}^{proj}/s,\mathbf{P}^*/s).\label{eq:poseloss}
\end{align}
The raw term supervises points before Kabsch. The projected term supervises $\widehat{\mathbf{P}}^{proj}=\widehat{\mathbf{R}}\mathbf{A}$ after complete rigid recovery, with gradients through both alignment and the inter-block pose updates.
Center supervision uses corrected centers. Training uses no separate quaternion loss.

Training samples $k\in\{1,\ldots,N\}$ and constructs the input at $t=(k-1)/N$, with $N=50$.
For dimensionless coordinates, the Gaussian loss is
\begin{equation}
 w_k=\frac{N(1-\sigma_1^{2/N})}{2\sigma_1^{2k/N}},\qquad
 \mathcal D_t(\widehat{\mathbf{P}},\mathbf{P})=\frac1M\sum_{m=1}^M
 \frac{w_{k_m}}{3|\mathcal A_m|}\sum_{u\in\mathcal A_m}\|\widehat{\mathbf{p}}_u-\mathbf{p}_u\|^2,
\end{equation}
where $\mathcal A_m$ contains valid anchors in MOF $m$.
Each MOF has equal weight. Time weights are not normalized across steps.
The rigidity term averages
$[(\|\widehat{\mathbf{p}}_{ik}^{raw}-\widehat{\mathbf{p}}_{il}^{raw}\|-\|\mathbf{a}_{ik}-\mathbf{a}_{il}\|)/s]^2$
over ordered valid within-block pairs, including the zero diagonal, with denominator $\sum_i K_i^2$. Coordinates inside this expression are in \AA.

The lattice state uses six entries
$(\log a,\log b,\log c,\tan(\alpha-\pi/2),\tan(\beta-\pi/2),\tan(\gamma-\pi/2))$,
with lengths in \AA, angles in radians, and no additional standardization.
It follows a Gaussian flow with $\sigma_1=0.0316228$. Its loss averages weighted squared error over six entries and then MOFs.
Each periodic center component follows MOF-BFN's circular flow, represented by a mean direction and concentration~\citep{jiao2025mofbfn}.

\subsection{Rigid pose recovery}
\label{app:solver}
Kabsch alignment in Equation~\ref{eq:anchorprojection} gives each block's initial rotation. The center head supplies its initial position.
To coordinate neighboring blocks, rigid recovery then applies two point-based pose updates, using Block GNN features and up to 16 fixed FPS samples $\mathbf{S}_i=\{\mathbf{s}_{iu}\}$ per block.
For each block pair, four learned queries attend to the local samples using element, coordinate, and radial features, conditioned on both block embeddings and the direction to the neighboring center.
Each query defines a local virtual point
\begin{equation}
 \mathbf{q}_{i\to j,m}=\sum_u\alpha_{ijmu}\mathbf{s}_{iu}+\ell_{ijm}\widehat{\mathbf{d}}_{ij}^{local},
 \qquad \sum_u\alpha_{ijmu}=1,\quad 0\leq\ell_{ijm}\leq4~\text{\AA}.
 \label{eq:interface}
\end{equation}
Here $\alpha_{ijmu}$ are attention weights and $\widehat{\mathbf{d}}_{ij}^{local}$ is the neighbor direction in block $i$'s frame.
These are learned meeting points. Each block has a separate set for each neighbor.

For edge $e=(i,j,m)$, let $\mathbf{u}_{ie}=\mathbf{R}_i\mathbf{q}_{i\to j,m}$ and $\mathbf{u}_{je}=\mathbf{R}_j\mathbf{q}_{j\to i,m}$.
The residual between paired world points is
\begin{equation}
 \mathbf{r}_e=\mathbf{c}_j+\mathbf{L}\mathbf{n}_{ij}+\mathbf{u}_{je}-\mathbf{c}_i-\mathbf{u}_{ie},\qquad
 \mathbf{n}_{ij}=-\operatorname{round}\big(\mathbf{L}^{-1}(\mathbf{c}_j-\mathbf{c}_i)\big).
 \label{eq:residual}
\end{equation}
Componentwise rounding selects a centered fractional-cell image. In a skew cell, this need not be the Euclidean nearest image.
A learned confidence $w_e$ weights each pair.
The translation drive sums $w_e\mathbf{r}_e$ at endpoint $i$ and $-w_e\mathbf{r}_e$ at $j$. The rotation drive sums $w_e(\mathbf{u}_{ie}\times \mathbf{r}_e)$ and $-w_e(\mathbf{u}_{je}\times \mathbf{r}_e)$, respectively.
We normalize translation by incident confidence and rotation by the confidence-weighted squared lever arms, with an additive stabilizer of one.
Learned scalar gates control both updates. Translation proposals are centered within each MOF and their norms capped at 0.35~\AA, while rotation increments are capped at 0.2 radians.
The poses update as
\begin{equation}
 \mathbf{c}_i\leftarrow \mathbf{c}_i+\delta \mathbf{c}_i,\qquad
 \mathbf{R}_i\leftarrow\exp([\boldsymbol{\delta\omega}_i]_\times)\mathbf{R}_i.
\end{equation}
Two iterations retain local virtual points and confidences while recomputing world offsets and periodic images.
The point features have width 64. The lattice is detached and guarded in these updates.
Supervision acts on the resulting centers and rigid anchors, with gradients through the complete recovery computation. No connection labels are needed.

\paragraph{Periodic-image convention.}
Equation~\ref{eq:residual} wraps each fractional displacement component into the centered unit interval.
For example, take lattice vectors $(10,0,0)$, $(9,3,0)$, and $(0,0,10)$~\AA\ and fractional displacement $(0.49,0.49,0)$.
Rounding chooses image $(0,0,0)$, giving distance $9.43$~\AA, whereas image $(0,-1,0)$ gives the nearest distance $1.56$~\AA.
On the full test set's reference cells and block centers, this wrapping differs from the nearest-image distance by more than $10^{-5}$~\AA\ for 71,875 of 463,494 unordered block pairs (15.51\%).
This measures the image convention on reference geometries, not its effect on generation accuracy.

\subsection{Training and generation procedure}
\label{app:algorithm}
Assembly training uses one sampled time per example.
The raw-coordinate loss teaches each anchor to reach its target position, while the inter-anchor distance loss limits distortion of the point prediction.
Supervision of the final rigid anchor coordinates trains the recovered block placement, with gradients through pose recovery (Equation~\ref{eq:poseloss}).

Generation uses the inherited \texttt{end\_back} sampler with $N=50$ prediction steps, starting from the state priors.
After each prediction, the recovered rotation acts on the fixed local geometry to give rigid anchor offsets $\widehat{\mathbf{P}}_i^{proj}=\widehat{\mathbf{R}}_i\mathbf{A}_i$.
These rigid anchors supply the clean estimate for the next generation step.
At step $n$, the network receives $\boldsymbol{\mu}_n$ at time $t=(n-1)/N$ and returns rigid anchors $\widehat{\mathbf{P}}_n^{proj}$. The sampler constructs
\begin{equation}
 \boldsymbol{\mu}_{n+1}=\gamma_{n/N}\widehat{\mathbf{P}}_n^{proj}/s+
 \sqrt{\gamma_{n/N}(1-\gamma_{n/N})}\,\boldsymbol{\epsilon}_n,
 \quad \boldsymbol{\epsilon}_n\sim\mathcal N(\mathbf{0},\mathbf{I}),\quad n<N.
 \label{eq:feedback}
\end{equation}
Lattice and center predictions likewise determine their next flow states.
Each call uses one native-scale sender draw.
The added point noise may again break rigidity, but each recovered block preserves its local geometry.
The final step reconstructs all atoms as $\widehat{\mathbf{y}}_{ia}=\widehat{\mathbf{c}}_i+\widehat{\mathbf{R}}_i \mathbf{x}_{ia}$.

\begin{table}[H]
\centering\small
\caption{\textbf{Training and generation with fixed fitted blocks.}}
\label{tab:algorithm}
\begin{tabular}{p{0.94\textwidth}}
\toprule
\textbf{Training with one sampled time per example}\\
1. Sample $k$ and draw lattice, center, and anchor states from their clean targets at $t=(k-1)/50$.\\
2. Encode atoms and blocks. Predict the lattice, centers, and raw clean anchors.\\
3. Recover rigid poses using Kabsch alignment and two inter-block point-based updates.\\
4. Reconstruct $\widehat{\mathbf{P}}^{proj}=\widehat{\mathbf{R}}\mathbf{A}$ and optimize Equation~\ref{eq:loss}.\\
\midrule
\textbf{Generation with fifty decoder calls}\\
1. Initialize lattice, periodic-center, and anchor states at their priors.\\
2. Predict clean lattice, corrected centers, and reconstructed anchors as above.\\
3. Unless this is the last call, draw the next native states from those clean predictions.\\
4. Return the final lattice and all atoms $\mathbf{y}_{ia}=\widehat{\mathbf{c}}_i+\widehat{\mathbf{R}}_i \mathbf{x}_{ia}$.\\
\bottomrule
\end{tabular}
\end{table}

\subsection{Data splits and evaluation}
\label{app:rms}
\label{app:selection}
We use AtomMOF's MOF-only benchmark and preprocessing, with 199,406 training, 24,926 validation, and 19,792 test structures.
The input supplies building blocks, not their crystal conformations.
The fitting model is validation-selected. Assembly and controlled comparisons use their final EMA checkpoints.
Full-test results cover all 19,792 structures.
Geometry diagnostics separately use 256 validation MOFs, with 128 validation MOFs for native sampling checks.
The anchor-coverage statistics in Appendix~\ref{app:anchorcoverage} use all test templates.

All-atom match rate evaluates the complete periodic structure.
We use \texttt{StructureMatcher} with site tolerance $0.5$ (loose) or $0.3$ (strict), length tolerance $0.3$, and angle tolerance $10^\circ$~\citep{ong2013pymatgen}.
Its reported RMS is dimensionless and defined as
\begin{equation}
 r_s=\frac{\sqrt{N_s^{-1}\sum_a\|\boldsymbol{\delta}_{sa}\|^2}}{(\overline V_s/N_s)^{1/3}},
 \qquad \overline r=\frac1{|\mathcal M|}\sum_{s\in\mathcal M}r_s,
\end{equation}
where $\mathcal M$ contains successful matches, $N_s$ is the site count after cell preprocessing, and $\overline V_s$ is the comparison-lattice volume.
Matching accounts for periodic images, compatible atom assignments, and common translation.
We average RMS over each method's successful subset and report Cartesian diagnostics elsewhere in \AA.

\label{app:sample5}
For five-candidate evaluation, seeds 123--127 produce 98,960 candidates.
Each run uses a different global random stream and a different anchor-specific random stream.
A structure matches if any candidate succeeds, and its RMS is the minimum over matching candidates.
Both tolerances use the same candidate bank and this best-of-five rule.
Baseline values in Table~\ref{tab:main} follow AtomMOF's published comparison.
\FloatBarrier

\FloatBarrier
\section{Supplementary experiments}
\label{app:posecontrols}
\label{app:components}
We examine the local geometry used for assembly, geometric inputs to the predictor, and the pose-state representation introduced in Sections~\ref{sec:geometry} and~\ref{sec:fit}.
We then report state diagnostics and additional architecture controls.
Match rates (MR) are percentages. Cartesian errors are in \AA.

\subsection{Effect of local conformation fitting}
\label{app:fitresults}
Table~\ref{tab:fit} evaluates the frozen fitting model on all held-out test blocks.
Mean aligned block RMSD decreases from 0.842 to 0.608~\AA.
Across 347,495 modeled torsions, mean absolute circular error decreases from $86.74^\circ$ to $60.57^\circ$.
Aligned RMSD improves for 79.82\% of flexible linkers and worsens for 20.18\%.
\begin{table}[H]
\centering\small
\caption{\textbf{Local fitting reduces rigid-aligned error on held-out blocks.} RMSD is in \AA. Each row averages over the listed population.}
\label{tab:fit}
\begin{tabular}{lrrrr}
\toprule
Population&Count&RDKit RMSD&Fitted RMSD&Reduction\\
\midrule
All blocks&123,861&0.842&0.608&27.86\%\\
All linkers&87,150&1.150&0.817&29.00\%\\
Flexible linkers&76,676&1.292&0.914&29.21\%\\
\bottomrule
\end{tabular}
\end{table}

To measure the effect on assembly, we train one model on fitted conformations and another on their original RDKit precursors.
Atom identities, crystal targets, and data splits remain fixed. Each conformation supplies its own FPS anchors and best-aligned target rotations.
Both models use the Gaussian-anchor assembly, the same rigid recovery, and the plateau schedule in Table~\ref{tab:settings}.
On all 19,792 test structures, fitted conformations raise loose MR from 34.00\% to 39.80\% and strict MR from 9.82\% to 12.36\%.
Reducing internal conformation error therefore improves end-to-end rigid assembly.

\paragraph{Reference-pose diagnostic.}
\label{app:fitreference}
We hold the fitted local conformations fixed and place every block at its reference center in the reference lattice.
For each block, we compute the full-atom proper Kabsch rotation against its crystal coordinates, reconstruct all atoms, and evaluate the resulting crystal with the \texttt{StructureMatcher} settings in Appendix~\ref{app:rms}.
Across all 19,792 test structures, this placement matches 16,960 structures (85.69\%) at $\stol=0.5$ and 9,603 (48.52\%) at $\stol=0.3$.
The rotations minimize per-block coordinate error, while the matcher evaluates complete crystals.

\subsection{Effect of explicit block geometry}
\label{app:geometryinputs}
We test whether the actual point layout adds information beyond the pooled Atom GNN feature.
Both variants use Bingham states and a current-relative rotation head. The control replaces the actual layout with size-matched generic points.
Actual geometry improves loose MR from 18.84\% to 32.19\%.

\label{app:genericframe}
The pair shares seed 123, batch 4,032, Bingham states, rotation supervision, and a current-relative quaternion head.
The replacement preserves point count and RMS radius but changes the object-specific layout.
For $K=1,2,3,4$, the unit-radius ordered frames are respectively
\begin{align}
 \mathbf{B}^{(1)}&=[(1,0,0)],\qquad \mathbf{B}^{(2)}=[(-1,0,0),(1,0,0)],\nonumber\\
 \mathbf{B}^{(3)}&=[(1,0,0),(-\tfrac12,\tfrac{\sqrt3}2,0),(-\tfrac12,-\tfrac{\sqrt3}2,0)],\nonumber\\
 \mathbf{B}^{(4)}&=\tfrac1{\sqrt3}[(1,1,1),(1,-1,-1),(-1,1,-1),(-1,-1,1)].
 \label{eq:genericframe}
\end{align}
We scale each frame by $r_i=\max(10^{-6},\sqrt{K_i^{-1}\sum_k\|\mathbf{a}_{ik}\|^2})$, measured about the full-block origin, without principal-component alignment.
Replacement affects local points, rotated points, periodic point features, and the pose descriptor and readout base derived from them.
Both variants retain the same point-order and anchor-kind embeddings. Full-block Atom GNN inputs, target rotations, and the Bingham process remain unchanged.
For degenerate point sets the derived pose is underdetermined. Restricting to the 2,627 test MOFs whose blocks all have four anchors still gives an actual-point advantage of 7.233/1.979 percentage points in loose/strict MR.

\subsection{Number of generated anchors}
\label{app:anchorcount}
We train complete Gaussian-anchor models with maximum FPS anchor counts of four, six, and eight.
The three runs retain fitted conformations, point decoding, Kabsch recovery, inter-block pose updates, and the same pose-loss weights.
All use 1,000 training epochs and effective batch size 4,032.
Table~\ref{tab:anchorcount} evaluates their final EMA models with 50 generation steps and one candidate on all 19,792 test structures.
We use four spatially dispersed anchors to keep the pose representation compact while retaining geometric coverage.
The fresh four-anchor run also supplies the reference in Table~\ref{tab:mainablation}. It is separate from the main benchmark checkpoint.

The comparison is reported in Table~\ref{tab:anchorcount} in Section~\ref{sec:components}.

\subsection{End-to-end pose-generation variants}
\label{app:matchedprotocol}
We evaluate four state--output combinations with fitted conformations and actual block geometry.
For each state type, Bingham rotations or Gaussian anchor coordinates, the decoder predicts either a rotation correction or anchor coordinates.
AnchorPose combines Gaussian anchor states with anchor-coordinate prediction.
All four retain the center and lattice branches and two inter-block pose updates.

\paragraph{Rotation-state anchor decoding.}
We retain MOF-BFN's quaternion Bingham flow, with 50 steps and terminal precision parameter 200.
The normalized eigenvector of the state matrix with the largest eigenvalue defines $\mathbf{q}_{i,t}$ and hence $\mathbf{R}_{i,t}$.
The Block GNN receives $\operatorname{vec}(\mathbf{q}_{i,t}\mathbf{q}_{i,t}^\top)$. The point decoder receives $\mathbf{R}_{i,t}\mathbf{A}_i/s$, local anchors, block features, and the same periodic position encoding as AnchorPose.
The decoder receives rendered points at their rigid scale, without the Gaussian state's shrinkage or radius-strength factor.
After point prediction, rigid recovery provides the clean quaternion used to simulate the Bingham state at the next time with the inherited endpoint-feedback sampler.
The model retains the three pose-loss terms and time weights in Appendix~\ref{app:config}.

Table~\ref{tab:mainablation} in Section~\ref{sec:mainablation} reports the complete-system results using final 1,000-epoch EMA weights, 50 generation steps, and one candidate.
AnchorPose has the highest match rates among these complete configurations.

\paragraph{Direct rotation outputs.}
We also train Bingham-state models that retain the fitted conformations and actual point inputs but output a rotation correction.
The Block GNN receives the current orientation descriptor. The decoder receives local anchors, their current rotated positions, block context, and periodic position encodings.
After processing point features, the head pools them within each block and predicts a quaternion correction relative to the current rotation.
Both models retain the center and lattice branches and two inter-block pose updates.

The two variants differ in supervision before these updates. One uses rotation loss, while the other compares anchor coordinates under the predicted and target rotations.
Both use the final anchor-coordinate loss after the updates, with outer pose weight four.
Table~\ref{tab:rotationoutput} reports final 1,000-epoch EMA models under the same full-test, 50-step, single-candidate protocol.
In this comparison, the rotation-output model with coordinate supervision achieves slightly higher match rates than the model with rotation supervision.
The coordinate-supervised variant also appears in Table~\ref{tab:mainablation}.

\begin{table}[htbp]
\centering\small
\caption{\textbf{Rotation-output models with explicit geometry.} Both use Bingham states, fitted blocks, actual point inputs, and final coordinate supervision. Match rates (\%) cover all 19,792 test structures.}
\label{tab:rotationoutput}
\begin{tabular}{lrr}
\toprule
Supervision before pose updates & Loose MR & Strict MR\\
\midrule
Rotation loss & 35.96 & 9.64\\
Anchor-coordinate loss & 36.41 & 9.86\\
\bottomrule
\end{tabular}
\end{table}

\paragraph{Gaussian-anchor state with direct rotation output.}
This variant retains the native Gaussian state and feeds its unprojected noisy anchor coordinates to the decoder.
It uses AnchorPose's radius-weighted orientation descriptor, with input Kabsch alignment supplying the current rotation $\mathbf{R}_{i,t}$.
The head averages encoded anchor features within each block and predicts a quaternion increment around identity, which defines a world-frame correction $\Delta \mathbf{R}_i$.
The pre-update rotation is $\widehat{\mathbf{R}}_i^{pre}=\Delta \mathbf{R}_i\mathbf{R}_{i,t}$. Two inter-block updates then give $\widehat{\mathbf{R}}_i^{post}$.
Kabsch therefore supplies the input rotation, while the output head directly predicts its correction.

We supervise both $\widehat{\mathbf{R}}_i^{pre}\mathbf{A}_i$ and $\widehat{\mathbf{R}}_i^{post}\mathbf{A}_i$ against $\mathbf{P}_i^*$, using the coordinate normalization and time weighting of $\mathcal D_t$ in Appendix~\ref{app:config}.
Each coordinate term has weight four in the assembly objective. This variant uses no separate rotation loss or free-point prediction loss.
Rigid outputs preserve inter-anchor distances, so we disable the identically zero rigidity penalty.
The final offsets $\widehat{\mathbf{R}}_i^{post}\mathbf{A}_i$ supply the clean estimate for the next Gaussian state, with fresh noise at the next time.
The center and lattice objectives remain unchanged.
Table~\ref{tab:mainablation} reports this model's final 1,000-epoch EMA checkpoint under the full-test, 50-step, single-candidate protocol.
The 500-epoch controls in Appendix~\ref{app:additionalcontrols} use separate models.

\paragraph{Geometry-dependent Bingham noise.}
We further test a Bingham-state variant whose rotation noise depends on the complete fitted block geometry.
It uses the full-atom tensor $\mathbf{J}_{\mathbf{X}_i}$ from Equation~\ref{eq:spatialmetric} to set direction-dependent quaternion concentrations.
Appendix~\ref{app:geometrybingham} defines the sender, posterior, precision schedule, and training-only calibration.
This changes the rotation-state process while retaining four-point coordinate prediction, Kabsch recovery, inter-block pose updates, and the coordinate-based pose objectives.
Table~\ref{tab:mainablation} in Section~\ref{sec:mainablation} compares this model with a fresh four-anchor Gaussian model after 1,000 epochs, using final EMA weights, 50 steps, and one candidate.
The anchor-coordinate configuration yields higher match rates in this complete-system comparison.

\paragraph{Inter-block pose updates.}
We also evaluate a model trained without the inter-block updates in Appendix~\ref{app:solver}, retaining fitted blocks, Gaussian anchor states, point prediction, Kabsch recovery, and the three pose-loss terms.
This variant trains for 1,000 epochs, with learning rate $10^{-4}$ during the final 500 epochs.
Table~\ref{tab:mainablation} in Section~\ref{sec:components} shows that Kabsch-only recovery retains substantial reconstruction performance, while the complete model achieves higher match rates.

\FloatBarrier
\subsection{Pose-state geometry across sizes and time}
\label{app:smallblocks}
\label{sec:diagnostics}
For fixed fitted coordinates $\mathbf{X}_i$ and target rotation $\mathbf{R}_i^*$, we measure
\begin{equation}
 \theta_i(\mathbf{R})=\arccos\left(\frac{\operatorname{tr}(\mathbf{R}_i^{*\top}\mathbf{R})-1}{2}\right),\qquad
 d_i(\mathbf{R})=\sqrt{\frac1{n_i}\sum_a\|\mathbf{R}\mathbf{x}_{ia}-\mathbf{R}_i^*\mathbf{x}_{ia}\|^2}.
 \label{eq:diagnosticmetrics}
\end{equation}
The common template and center isolate orientation from conformation, lattice, and translation errors.
We measure these errors at the noisy input, the Kabsch estimate, and the final recovered pose.

The probe contains 256 validation MOFs and 1,838 blocks.
We retain 1,312 blocks with four valid, observable anchors, excluding 526 blocks (28.6\%) with insufficient or degenerate anchors.
We use three pose-noise draws at steps 20, 30, 40, 45, and 50.
Targets, fitted blocks, centers, and lattice-noise inputs are paired. Each pose state follows its native noise schedule.
For post-prediction and rollout measurements, we use final 1,000-epoch EMA models with the same point decoder, raw/rigidity/projected losses, and the same rigid recovery.
Each checkpoint is evaluated with three noise draws per block. This probe measures within-model noise variation.
The end-to-end pose-generation variants appear in Table~\ref{tab:mainablation}.

We divide blocks by full-atom RMS radius into groups of 438, 437, and 437, with cutoffs 2.880 and 3.784~\AA.
At step 50, Gaussian-anchor median input angles are $12.62/8.05/6.39^\circ$, compared with Bingham's $12.95/13.17/13.25^\circ$.
The corresponding atomic displacements are $0.367/0.344/0.347$~\AA\ and $0.394/0.604/0.869$~\AA.
Figure~\ref{fig:diagnostics} in Section~\ref{sec:main_diagnostics} combines the size dependence with its evolution across time.
At step 50, Bingham angular error is similar across sizes, while atomic displacement grows with block radius.
Gaussian-anchor angular error decreases with size, keeping atomic displacement near 0.35~\AA.
This is consistent with the geometric dependence in Section~\ref{sec:geometry}.
Across time, the two native schedules yield lower early errors for Bingham and lower late errors for Gaussian anchors.
Table~\ref{tab:stage} additionally compares the input, Kabsch estimate, and final recovered pose.

\begin{table}[htbp]
\centering\small
\caption{\textbf{Spatial pose error from input to recovered pose.} Median full-template RMS displacement in \AA, relative to the same target rigid pose.}
\label{tab:stage}
\begin{tabular}{rlrrr}
\toprule
Step&State / point decoder&Input&Kabsch estimate&Recovered pose\\
\midrule
20&Gaussian anchors&2.047&0.935&0.881\\
20&Bingham rendered points&0.957&0.515&0.438\\
30&Gaussian anchors&1.182&0.555&0.517\\
30&Bingham rendered points&0.786&0.429&0.361\\
40&Gaussian anchors&0.635&0.335&0.302\\
40&Bingham rendered points&0.674&0.399&0.331\\
50&Gaussian anchors&0.353&0.223&0.203\\
50&Bingham rendered points&0.581&0.386&0.307\\
\bottomrule
\end{tabular}
\end{table}

\paragraph{Directional balance within small blocks.}
The small-block group contains 438 blocks from 140 MOFs.
We measure anchor balance by $\eta=\lambda_{\min}(\mathbf{J}_{\mathbf{A}})/\operatorname{tr}(\mathbf{J}_{\mathbf{A}})$, which quantifies relative rotational constraint along the weakest axis.
Cutoffs 0.12748 and 0.21727 give three groups of 146 blocks.
Table~\ref{tab:direction} averages each block's three draws before computing group means.

\begin{table}[H]
\centering\small
\caption{\textbf{Step-50 angular error by anchor balance.} Means in degrees. Bingham has lower mean final error in the low-balance group. Gaussian anchors have lower means in the other two groups.}
\label{tab:direction}
\begin{tabular}{lrrrrr}
\toprule
&\multicolumn{2}{c}{Gaussian anchors}&\multicolumn{2}{c}{Bingham points}&\\
Anchor balance&Input&Final&Input&Final&Final difference$^\dagger$\\
\midrule
Low&17.80&10.80&13.76&9.87&$-0.93$\\
Medium&11.29&6.54&13.48&9.46&$+2.92$\\
High&14.90&6.38&13.49&12.28&$+5.89$\\
\bottomrule
\multicolumn{6}{l}{\footnotesize $^\dagger$Bingham minus Gaussian.}\\
\end{tabular}
\end{table}
The median paired differences favor Gaussian anchors in all three groups, with values of $+0.82/+2.07/+2.87^\circ$.
In the low-balance group, the opposite sign of the mean reflects sensitivity to large paired errors. Bingham's lower mean does not imply an advantage for the typical block.
Sizes and chemical types also vary across these groups. The comparison measures an association with geometry rather than an isolated geometry intervention.

\paragraph{Native sampling transitions.}
A separate 50-step rollout on 128 validation MOFs measures spatial displacement from the previous clean prediction to the next corrupted input, without target alignment.
At step 50, the median is 0.358~\AA\ for Gaussian anchors and 0.589~\AA\ for Bingham.
Small/medium/large values are $0.363/0.366/0.349$~\AA\ and $0.378/0.634/0.887$~\AA, respectively, reproducing the size-dependent pattern during sampling.
Consecutive clean predictions change by 0.174 and 0.157~\AA. Injected noise and output stability are distinct quantities.

\FloatBarrier
\subsection{Additional architecture and supervision controls}
\label{app:additionalcontrols}
Table~\ref{tab:controlmap} examines state rendering, anchor attention, pose readout, and supervision in smaller paired studies.
Within each pair, the architecture and training settings are shared except for the stated factor.

\begin{table}[htbp]
\centering\small
\caption{\textbf{Additional architecture and supervision controls.} Each group is a separate paired comparison, using final 500-epoch EMA weights, 50 steps, one candidate, and all 19,792 test structures.}
\label{tab:controlmap}
\begin{tabular}{p{0.70\textwidth}rr}
\toprule
Variant&Loose MR&Strict MR\\
\midrule
\multicolumn{3}{l}{\emph{State process with actual points and a relative rotation head}}\\
Bingham state, rigidly rendered points&32.185&7.801\\
Gaussian state, rigidly projected points&29.815&6.796\\
\midrule
\multicolumn{3}{l}{\emph{Anchor attention with a Gaussian state and point head}}\\
Within-block anchor attention&36.116&10.459\\
MOF-wide anchor attention&36.045&10.272\\
\midrule
\multicolumn{3}{l}{\emph{Readout with a Gaussian state and projected all-atom loss}}\\
Clean points followed by Kabsch&32.357&7.842\\
Current-relative rotation&32.609&7.917\\
\midrule
\multicolumn{3}{l}{\emph{Supervision with a Gaussian state and relative rotation head}}\\
Rotation loss&30.098&6.482\\
Projected all-atom coordinate loss&27.840&5.644\\
\bottomrule
\end{tabular}
\end{table}

The rendered-state pair supplies rigid current-pose points to a relative rotation head with rotation supervision.
Bingham achieves higher MR in this setting (32.185\% versus 29.815\% loose MR).
The attention pair retains inter-block Block GNN messages and uses Kabsch-only pose recovery in both variants. Within-block and MOF-wide anchor attention give similar results in this comparison.

For the readout pair, both variants use Gaussian states and projected all-atom supervision, without inter-block pose updates or raw/rigidity losses.
In this readout comparison, point prediction followed by Kabsch and current-relative rotation attain similar loose MR (32.357\% and 32.609\%).
The supervision pair fixes the relative rotation head. Rotation loss gives 30.098\% loose MR versus 27.840\% for projected coordinate loss.
The relative performance of state representations, readouts, and objectives varies across these tested configurations.
\FloatBarrier

\FloatBarrier
\section{Qualitative examples}
\label{app:examples}

\subsection{Representative atoms and complete blocks}
\begin{figure}[htbp]
\centering
\includegraphics[width=0.94\textwidth]{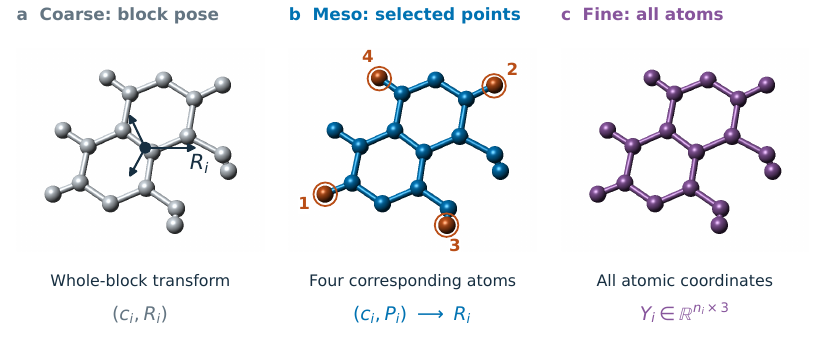}
\caption{\textbf{Three modeling resolutions for the same block.} Block pose, selected corresponding atoms, and all atomic coordinates define different generation variables. AnchorPose predicts the orange-ringed atoms and reconstructs the complete block by rigid alignment. Panels share camera and scale.}
\label{fig:representation}
\end{figure}

Figure~\ref{fig:representation} uses the fitted 18-atom linker from test structure 14094, block 3, selected by shape statistics rather than prediction accuracy.
Panel (a) of Figure~\ref{fig:motivation} compares the compact 18-atom linker with a nearly linear 12-atom linker from test structure 19284, block 4, in one shared three-dimensional scene. Their centers are offset along a common rotation axis for display. This arrangement is not a crystal configuration. Both are rotated by $20^\circ$ about their third principal axis through the centroid, giving full-atom RMS displacements of 0.91 and 1.42~\AA, respectively.
Panels (b) and (c) use a nearly linear 12-atom linker from test structure 19284, block 4, whose long and transverse rotation axes are easy to distinguish.
Panel (b) uses two atoms of the same 12-atom linker, at perpendicular distances of 0.63 and 5.55~\AA\ from the transverse rotation axis. Their displacements are 0.22 and 1.93~\AA, respectively.
The rotations preserve all interatomic distances, and the displayed RMS displacements are computed from every atom before and after rotation without subsequent alignment.
Figure~\ref{fig:fitposeerrors} uses saved fitted and crystal coordinates from test structure 344, block 3. An illustrative $12^\circ$ rotation separates pose error from the remaining internal conformation error.

Figure~\ref{fig:architecture} uses test structure 93, with its 26-atom block 2 highlighted within the 76-atom unit cell. Periodic neighbors complete blocks and coordination connections across the cell boundary. The noisy anchor states illustrate the generation variables.

\subsection{Pose-prediction examples}
\label{app:poseexamples}
\begin{figure}[htbp]
\centering
\includegraphics[width=0.94\textwidth]{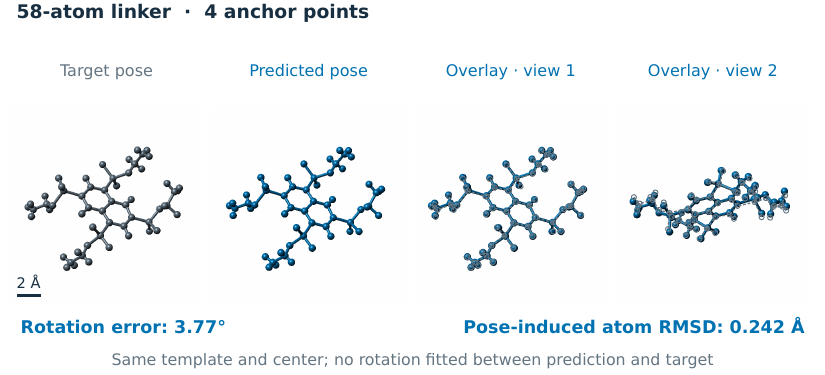}
\caption{\textbf{Accurate linker orientation in a generated structure.} Predicted and target rotations place the same 58-atom fitted template. Blue shows the prediction and dashed gray shows the target.}
\label{fig:reconstruction}
\end{figure}

\begin{figure}[htbp]
\centering
\includegraphics[width=0.94\textwidth]{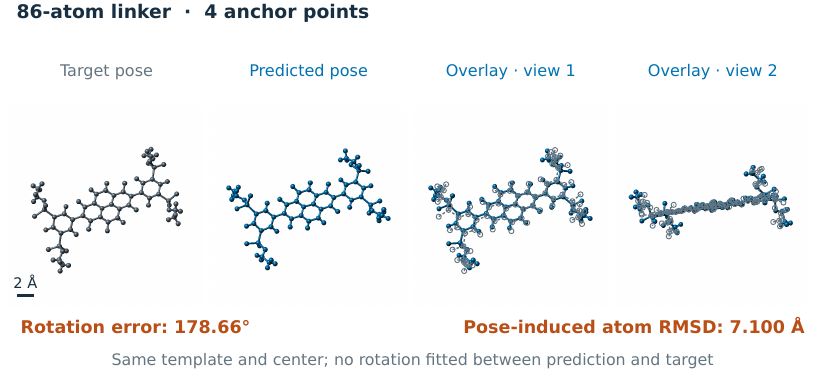}
\caption{\textbf{A pose error in the same generated structure.} The accompanying 86-atom linker illustrates a failure under the same fixed-correspondence, pose-only evaluation.}
\label{fig:posefailure}
\end{figure}

Figures~\ref{fig:reconstruction} and~\ref{fig:posefailure} show both nonsingleton blocks from one saved test-39 draw, generated with final EMA weights and 50 steps in CPU float32, separately from the official candidate bank.
Each overlay compares $\widehat{\mathbf{R}}_i\mathbf{X}_i$ with $\mathbf{R}_i^*\mathbf{X}_i$ at the same center, isolating orientation error.
Overlays share cameras and scales, without extra rotational alignment or atom permutation.
The 58-atom linker has error $3.768^\circ$ and pose RMSD 0.242~\AA. The 86-atom linker has error $178.660^\circ$ and RMSD 7.100~\AA.
Errors use the fixed atom correspondences of the assembly targets.

\FloatBarrier
\subsection{Local-conformation examples}
\label{app:localexamples}
\begin{figure}[htbp]
\centering
\includegraphics[width=0.88\textwidth]{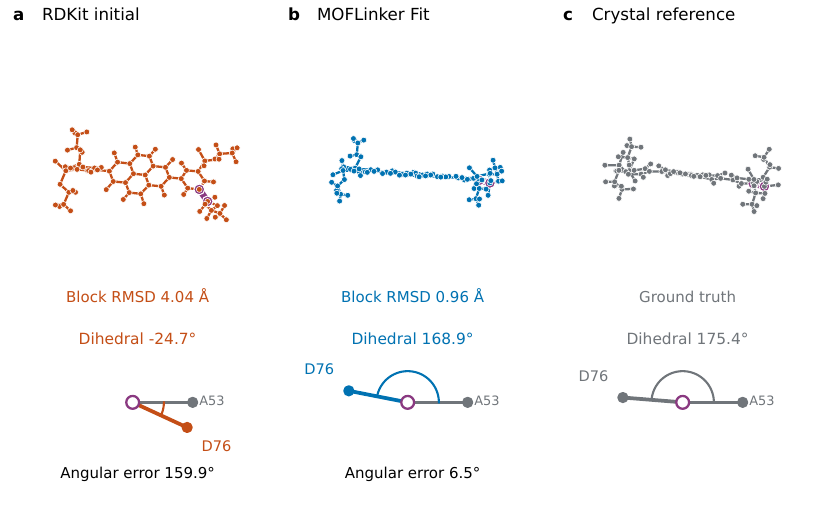}
\caption{\textbf{Local fitting reduces a large conformational discrepancy.} We rigidly align initial, fitted, and crystal coordinates to a common reference. Angular diagrams show measured dihedrals.}
\label{fig:fitexamples}
\end{figure}

The examples come from 6,337 blocks in the first 1,024 test structures.
The improvement example is drawn from the two largest improvements in distinct MOFs among 24--90-atom blocks with initial aligned RMSD above 2~\AA\ and fitted RMSD below 1~\AA.
Figure~\ref{fig:fitexamples} shows test structure 39, block 3, where RMSD decreases from 4.037 to 0.964~\AA.
The highlighted dihedral has initial, fitted, and crystal angles $-24.74^\circ$, $168.91^\circ$, and $175.38^\circ$.
Fitting includes both torsion and coordinate updates.

\begin{figure}[htbp]
\centering
\includegraphics[width=0.88\textwidth]{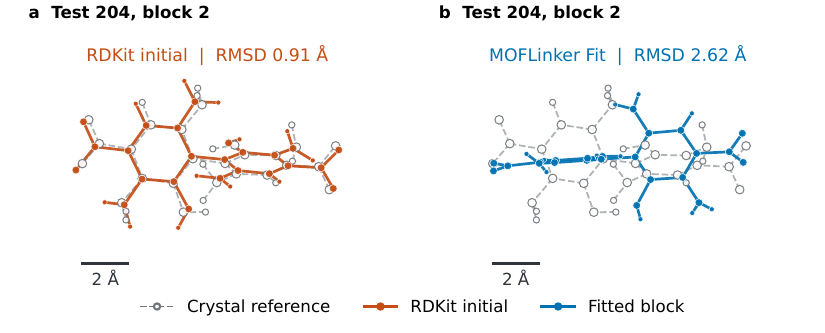}
\caption{\textbf{A failure of local fitting.} Rigid assembly alone cannot correct an internal discrepancy remaining in the fitted template.}
\label{fig:fitfailure}
\end{figure}

Figure~\ref{fig:fitfailure} is the largest worsening among 24--90-atom linkers with initial RMSD above 0.2~\AA.
The full-test frequency of worsening is 20.18\% among flexible linkers (Table~\ref{tab:fit}).
All conformation comparisons remove rigid pose using full-atom alignment.

\FloatBarrier

\end{document}